\documentclass{article}

\usepackage{PRIMEarxiv}

\usepackage[utf8]{inputenc} 
\usepackage[T1]{fontenc}    
\usepackage{hyperref}       
\usepackage{url}            
\usepackage{booktabs}       
\usepackage{amsfonts}       
\usepackage{nicefrac}       
\usepackage{microtype}      
\usepackage{lipsum}
\usepackage{fancyhdr}       
\usepackage{graphicx}       
\usepackage{multirow}
\usepackage{longtable}
\usepackage{booktabs}
\usepackage{amssymb}
\graphicspath{{media/}}     
\usepackage{float}

\pdfoutput=1

\title{Towards the Human Digital Twin: Definition and Design – A survey
}

\author{
  Martin Wolfgang Lauer-Schmaltz \\
  Department of Technology, Management and Economics \\
  Technical University of Denmark \\
  2800 Kgs. Lyngby, Denmark\\
  \texttt{mwola@dtu.dk} \\
   \And
  Philip Cash \\
  Northumbria School of Design \\
  Northumbria University \\
  Newcastle upon Tyne NE1 8ST, United Kingdom\\
  \texttt{philip.cash@northumbria.ac.uk} \\
  \And
  John Paulin Hansen \\
  Department of Technology, Management and Economics \\
  Technical University of Denmark \\
  2800 Kgs. Lyngby, Denmark\\
  \texttt{jpha@dtu.dk} \\
  \And
  Anja Maier \\
  Department of Technology, Management and Economics \\
  Technical University of Denmark \\
  2800 Kgs. Lyngby, Denmark\\
  \\
  Department of Design\\ 
  Manufacturing and Engineering Management\\ 
  University of Strathclyde, Glasgow G1 1XH, United Kingdom\\
  \texttt{anja.maier@strath.ac.uk} \\
}

\begin{document}
\maketitle

\begin{abstract}
Human Digital Twins (HDTs) are a fast-emerging technology with significant potential in fields ranging from healthcare to sports. HDTs extend the traditional understanding of Digital Twins by representing humans as the underlying physical entity. This has introduced several significant challenges, including ambiguity in the definition of HDTs and a lack of guidance for their design. This survey brings together the recent advances in the field of HDTs to guide future developers by proposing a first cross-domain definition of HDTs based on their characteristics, as well as eleven key design considerations that emerge from the associated challenges.
\end{abstract}


\section{Introduction}
\label{Section1}
Digital Twins (DTs) are a critical technology for digitalizing physical entities in domains ranging from industry to city planning \cite{barricelli2019, siqueira2021}. DTs’ ability to continuously adapt to a physical entity’s state, simulate future events, and actively influence feedback and decision processes, goes significantly beyond traditional digital models as merely representations \cite{singh2021}.  Thus, Industry 4.0 has started using DTs—along with other cutting-edge technologies, such as the Internet of Things (IoT), Big Data, and Artificial Intelligence (AI)—to significantly increase the efficiency and safety of both products and processes \cite{singh2021}. Further, due to DTs’ real-time monitoring and simulation capabilities, they are being increasingly adapted to domains such as healthcare to meet demands for individualized diagnostics and treatment \cite{boulos2021}. For example, DTs have been used to improve clinical decision processes by representing human individuals, i.e., Human Digital Twins (HDTs) \cite{rivera2019}. However, the increased focus on human factors in HDT research introduces substantial new challenges to the traditional understanding of DTs.\\
Despite the increasing interest in HDTs in the DT literature, no consensus exists on a cross-domain definition of HDTs and how they differ from traditional DTs. Instead, most research on HDTs either limits their characterization to particular use cases or discusses HDTs as simply another application of DTs. However, humans differ considerably from the physical entities traditionally represented by DTs, especially given humans’ individuality and the involvement of non-physical factors, such as cognitive mechanisms. Thus, there is a critical need to clarify how HDTs challenge traditional understandings of DTs.\\
In response to this need, this paper conducts a systematic literature review to distill essential characteristics, emerging challenges, and key design considerations for HDTs, as a foundation for creating a first cross-domain understanding of the concept. The main contributions of this review thus lie in answering the following research questions:
\begin{itemize}
\item How can HDTs be defined and conceptualized in a general, cross-domain context?
\item What categories of HDTs emerge from the HDT literature, and how do they differ from traditional DTs? 
\item What significant challenges arise from the differences between HDTs and traditional DTs?
\item What HDT design considerations result from these challenges?
\end{itemize}

\section{Digital Twin}
\label{Section2}
The idea of using digital representations of physical entities to monitor and simulate their internal properties dates back to well before the introduction of the term “Digital Twin”. In 1960s, NASA used simulators to replicate and simulate flight conditions of a spacecraft that ultimately lead to the success of the Apollo 13 mission in 1970 \cite{allen2021}. In the early 1990s, Gelernter introduced the concept of “mirror worlds”, that is, the idea to virtually mirror reality to perceive the world in a more detailed way \cite{gelernter1993}. However, it was not until the early 2000s that this idea was developed into a stand-alone concept, resulting in what is today known as the DT.
For clarity about the origin of HDTs and their differentiation from this traditional DT concept, this section gives a brief overview of DTs, their application fields nowadays, and enabling architecture, building on prior reviews in this area \cite{barricelli2019, Fuller2020, hinduja2020, jacoby2020}.

\subsection{Definition and Characteristics}
\label{Section2.1}
\begin{figure} [H]
  \centering
  \includegraphics[scale=0.8]{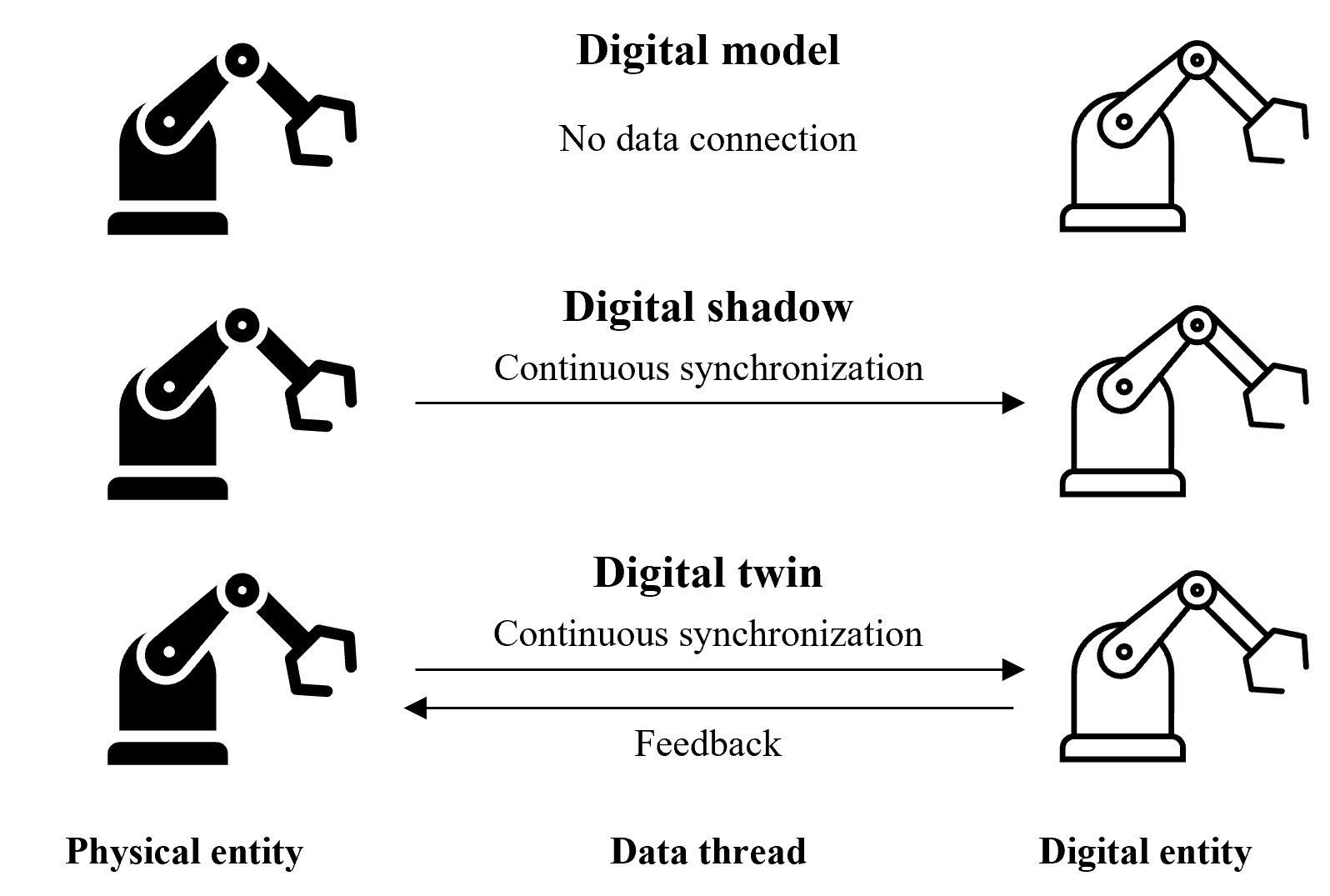}
  \caption{Comparison of the three digital concepts for digitally representing a physical entity. The concepts differ in the development of the connecting data thread.}
  \label{Fig.1}
\end{figure}

In 2003, Grieves introduced the concept of DTs as a “virtual representation of what has been produced” \cite{grieves2015}. The first mention of DTs by name, however, was in a draft of NASA’s technological roadmap in 2010, where the concept was explored primarily for (air) vehicles as “an integrated multi-physics, multi-scale, probabilistic simulation of a vehicle or system that uses the best available physical models, sensor updates, fleet history, etc., to mirror the life of its flying twin.” \cite{shafto2010}. Current literature generalizes this definition to DTs being the continuous and accurate digital representation of a physical object, process, or system that allows accessing, monitoring, and optimizing their state using simulation and prediction methods \cite{barricelli2019, Angulo2020}. As such, a DT usually consists of three components: the physical entity to be represented, a digital entity, and a bi-directional data thread that allows (continuous) synchronization between the digital and physical entity, as well as feedback by the digital entity to the physical entity \cite{barricelli2019}.\\
However, a notable inconsistency exists between the definition and applied concepts in the current DT literature. For example, some authors refer to systems as DTs that are a digital representation of the physical entity yet do not allow bi-directional data exchange. Although a clear delineation is often tricky, such systems are referred to as “digital shadows” or “digital models” \cite{Fuller2020}. In contrast to DTs, digital shadows refer to digital representations of physical entities connected to the physical entity via a unidirectional data thread to (continuously) adapt to the physical entity’s current state without influencing it through feedback \cite{Fuller2020}. Digital models refer to digital representations of physical entities not connected to the latter and thus neither (continuously) adapt to the physical entity’s current state nor influence it through feedback \cite{Fuller2020}. As such, there is a significant difference between the three digital concepts, particularly regarding their degree of interaction with the physical entity. Figure \ref{Fig.1} visualizes these concepts and their differences in relation to the physical entity.
\newpage
In addition to this definition of DTs, there exist several widely recognized characteristics that can be summarized as follows \cite{barricelli2019, singh2021, Fuller2020}: 
\begin{itemize}
\item \textbf{Unique identifiability:} DTs are assignable to their physical entity to avoid confusion.
\item \textbf{Fidelity:} DTs represent the physical entity with sufficient accuracy, given the specific use case.
\item \textbf{Continuous data exchange:} DTs reflect the physical entity’s true state at all times and can affect the physical entity with feedback.
\item \textbf{Data processing:} DTs aggregate raw data and predict future events to provide feedback to the physical entity or user. 
\item \textbf{Integrability:} DTs can communicate and collaborate with other DTs in the same DT system.
\item \textbf{Adaptability:} DTs represent the current state of the physical entity at all times and adjust to changes in use case and environment.
\end{itemize}

\subsubsection{Application Fields}
\label{Section2.2}
Given the characteristics outlined in the previous section, DTs have been applied across domains. However, different use-cases pose different requirements to the DT, resulting in variation in DT concepts across domains. Thus, to better understand this use-case-based variation, we examine a number of typical application examples from across domains, drawing on recent reviews (i. e., \cite{barricelli2019, Fuller2020}).
\begin{itemize}
\item \textbf{Industrial applications:} Apart from exceptions such as supply-chain DTs or DT-driven fully-automated factories, DTs primarily represent specific products, as well as physical processes and systems. In the automotive industry, for instance, vehicle DTs allow monitoring of the vehicle’s state and interaction between user and vehicle \cite{singh2021}. In the manufacturing industry, DTs of robotic cells or production lines are often used to improve the efficiency and safety of the production process \cite{bilberg2019, perez2020}. DTs in industrial applications thus mainly exist as product avatars or as the digital representation of physical processes consisting of multiple physical entities.
\item \textbf{Energy and Smart Cities:} In addition to representing physical objects, such as wind turbines (e. g., \cite{fahim2020}) or physical processes and systems such as whole power plants (e. g., \cite{lei2022}), DTs increasingly represent large-scale systems that include non-physical processes and factors. For example, DTs of energy grids allow monitoring and prediction of fluctuations in the energy demand and consumption behavior for optimal energy distribution \cite{you2022}. Furthermore, DTs of cities or specific districts allow surveilling and optimizing the city’s infrastructure based on public commute behavior as well as supporting urban planning by predicting ecological and ethical effects \cite{shirowzhan2020, white2021}. These use-cases reveal a shift within the conceptualization of DTs from representing solely physical systems to more data-driven approaches that also consider non-physical factors.
\item \textbf{E-Commerce:} Apart from representing particular products or optimizing the supply chain and other infrastructural factors, DTs have recently been used to represent other involved parties, such as human customers. Here, user HDTs allow, for example, tracking the user’s shopping behavior (e.g., their search history, past purchases, and ratings) and thus play an increasing role in recommender systems \cite{vijayakumar2020}. As such, user HDTs in E-Commerce applications do not represent a technological product or process but entail data-based models of behavioral factors (e.g., site navigation and purchase decisions). 
\item \textbf{Healthcare:} Like other domains, DTs in healthcare mainly represent particular objects (e.g., medical equipment), (physical) processes (e. g., infrastructure or patient pathways), and patients \cite{barricelli2019}. However, patient HDTs are not limited to behavioral factors and can also represent specific body parts, organs, or physiological factors \cite{boulos2021}. Thus, certain applications require the DT concepts to go beyond the differentiation between system-based and data-driven approaches and instead combine physical and non-physical factors. 
\end{itemize}
Cutting across the different domains, DTs show common characteristics regarding their architecture and enabling technologies. However, the above examples also reveal how new types of DTs introduce new considerations. In particular, the emerging concept of HDTs poses significant challenges arising from the inclusion of human factors and associated new design considerations. Before we discuss these HDT-specific issues, we first provide an overview of common architecture and enabling technologies.

\subsubsection{Architecture and Enabling Technologies}
\label{Section2.3}
\begin{figure} [H]
  \centering
  \includegraphics[scale=0.6]{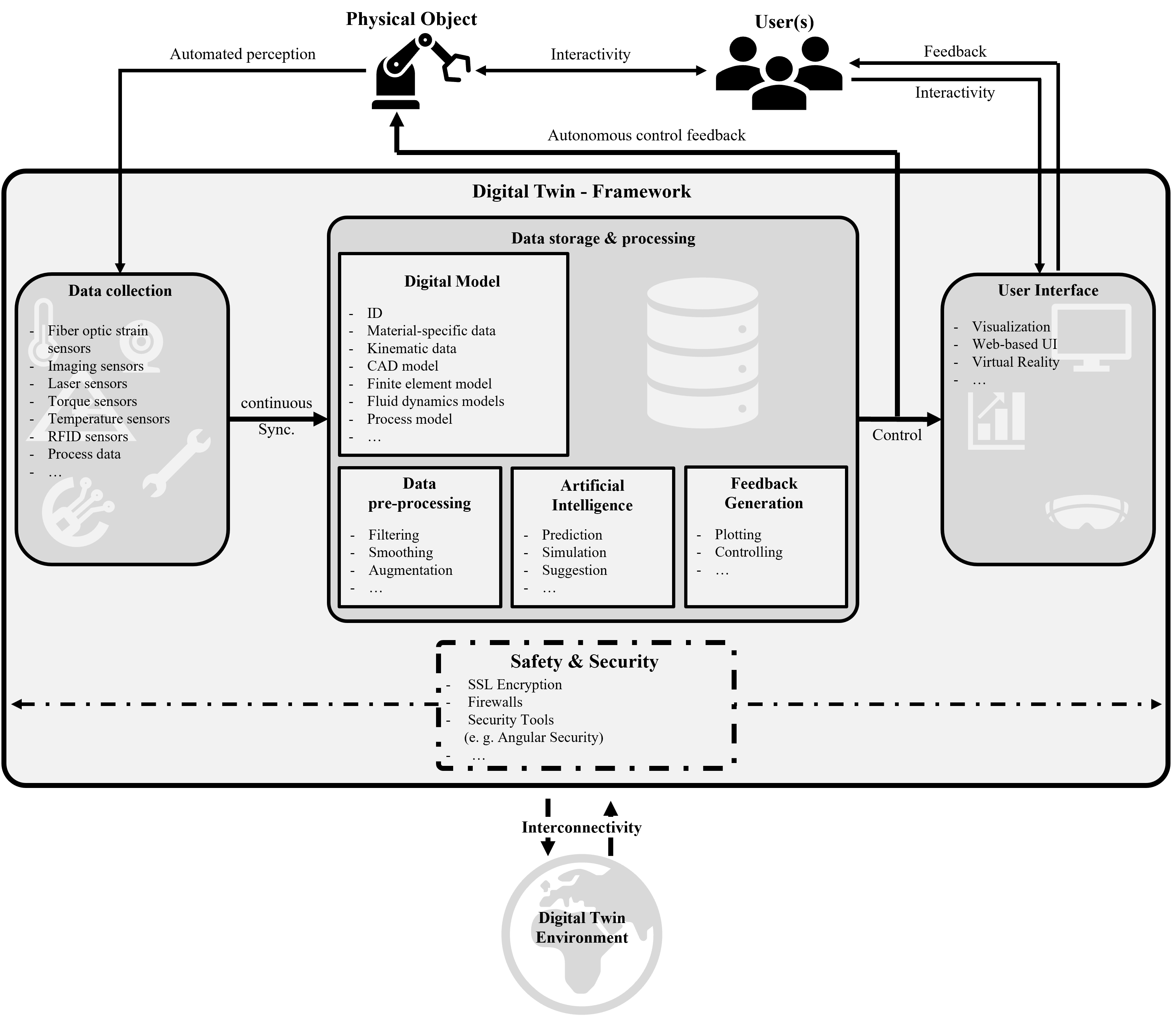}
  \caption{Typical Digital Twin framework in conventional application domains. The physical object is monitored using sensors, a continuously updated digital representation is stored using a database, simulations based on Artificial Intelligence are performed and corresponding feedback is provided to the physical world.}
  \label{Fig.2}
\end{figure}

To establish how HDTs build on traditional DTs and where they differ, this section briefly describes the common architecture and enabling technologies of traditional DTs. Here, a DT typically consists of three layers (see Figure \ref{Fig.2}).\\
First, a data collection layer, with integrated sensors, is used to perceive the physical entity’s current state \cite{Fuller2020, Mohapatra2020}. The specific type of sensor technology varies depending on the application. For example, in the aerospace domain, fiber optic strain sensors and imaging sensors can be used to monitor the aircraft’s state \cite{milanoski2021}, whereas, in the energy sector, temperature sensors are required to monitor the thermal properties of power plants \cite{junior2021}. Despite this variation, sensors are typically integrated directly into the physical entity.\\
Second, a data storage and processing layer is used to aggregate and store the collected raw data and to generate feedback based on prediction and simulation \cite{Fuller2020, Mohapatra2020}. Due to their ability to simplify the handling of even massive data packs, various database programs such as MongoDB and MySQL are commonly used in recent DT approaches \cite{Fuller2020}. In addition, various DTs extend the stored data to models using modeling tools. Computer Aided Design (CAD), for instance, can provide three-dimensional virtual models with defined material properties in mechanical applications \cite{sierla2020}, whereas computational fluid dynamics modeling allows for modeling fluid flows \cite{molinaro2021}. As such, models organize the data stored in databases to accurately represent the physical entity and its relevant properties.\\
Further, DTs use these data and models to predict the physical entity’s future states and generate corresponding feedback. Particularly the integration of AI allows for accurate and robust decision-making. For example, predictive deep learning models can estimate specific states based on the available data, such as the remaining lifetime of products or materials (e.g., \cite{han2021}) or specific health parameters in human-centered healthcare \cite{rathmore2021, shamanna2020_1}. Alternatively, reinforcement learning allows the direct mapping of DT data to intelligent feedback by learning optimal control schemes using outcome-based reward systems \cite{rathmore2021}. Thus, the integration of AI allows for intelligent and robust data aggregation and feedback generation.\\
Third, in contrast to digital models or digital shadows, DTs can directly influence the physical entity via an output and interaction layer \cite{Fuller2020, Mohapatra2020}. This can provide autonomous feedback as well as decision support to the relevant users, depending on the application. To directly interact with the physical entity, DTs provide a data thread that links the data processing unit with the physical entity. For this purpose, commonly used connection technologies are optical, coaxial, or twisted pair cables, and wireless methods such as Zig-Bee, Bluetooth, Wi-fi, ultra-wideband, or NFC \cite{hu2021}. To communicate with human users, common examples of user interface technology are web-based applications (e.g., \cite{han2021}) or virtual/augmented reality systems (e.g., \cite{zhu20219}). While web-based applications are compatible with a broader range of devices \cite{han2021}, virtual/augmented reality allows for immersion and more intuitive data visualization, for example, by visualizing important information directly on the observed entity \cite{zhu20219, bravo2020}. Thus, providing intelligent feedback is a key capability of DTs to allow for dynamic system response or interaction with human users.\\
Finally, security plays a critical cross-layer role in the application of DTs. The large, centralized, and highly specific amounts of sensitive data make DTs vulnerable to cyber-attacks \cite{barricelli2019}. However, several well-developed and standardized security measures for data transmission and storage exist, such as data encryption using DES/AES algorithms, authentication measures, and state-of-the-art blockchain technologies \cite{mashaly2021}. Thus, despite a potentially negative effect on the DT’s overall performance, their combination provides security across all layers. 
\newpage

\section{Methodology}
\label{Section3}
\begin{figure}[H]
  \centering
  \includegraphics[scale=0.6]{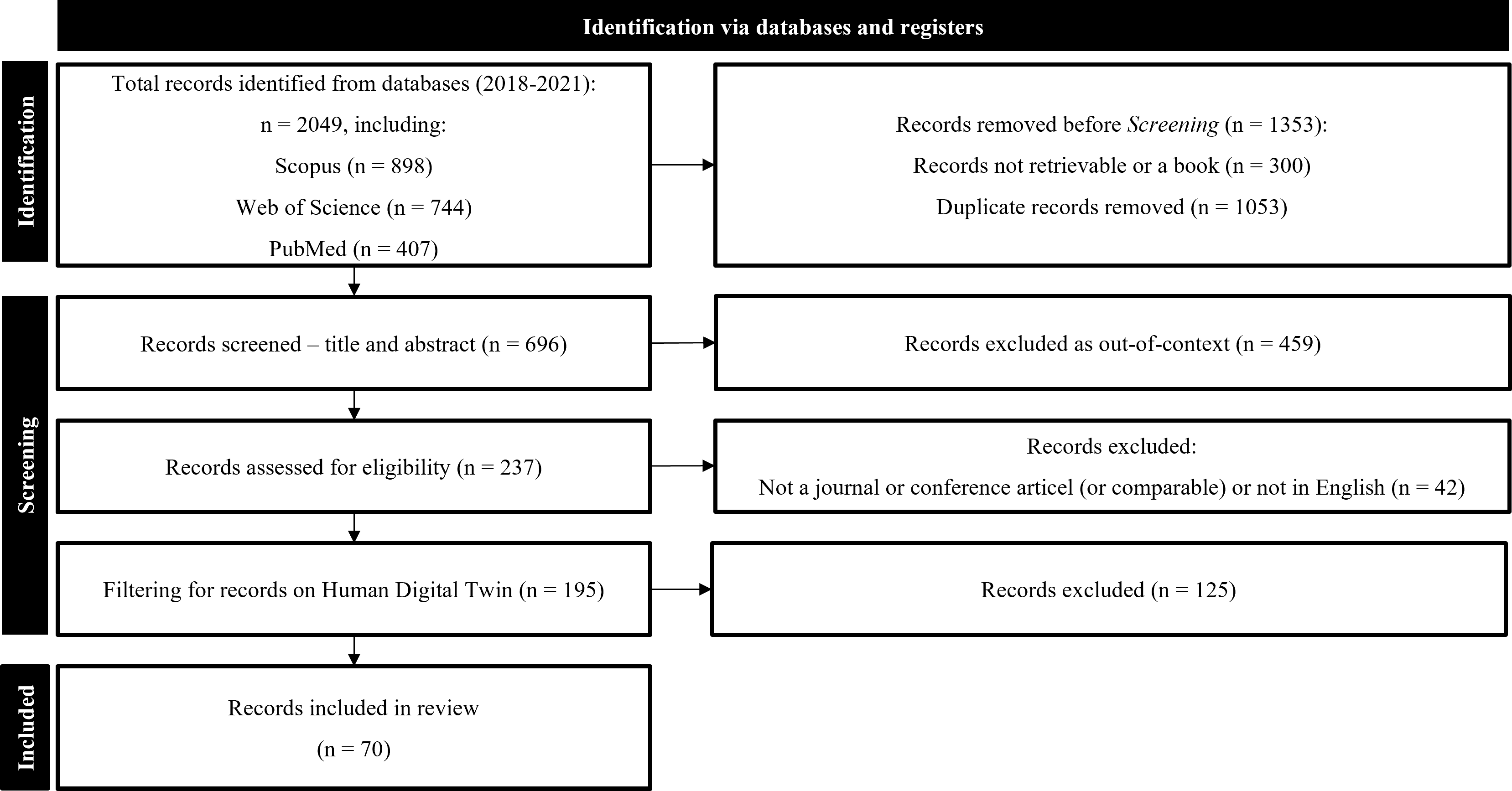}
  \caption{PRISMA diagram summarising the systematic literature review of Human Digital Twins.}
  \label{Fig.3}
\end{figure}

To answer our four research questions, we conducted a systematic literature review. The review followed the updated Preferred Reporting Items for Systematic Reviews and Meta-Analyses (PRISMA) statement consisting of three main phases (1) Identification, (2) Screening, and (3) Inclusion \cite{page2021}. Figure \ref{Fig.3} gives a detailed overview of the approach.
\newpage
To bring together insights from cross-domain literature, we conducted the search using three major databases: Scopus, Web of Science, and PubMED. While Scopus and Web of Science are prominent multi-disciplinary databases, PubMed primarily represents biomedicine and health-specific topics. Since this paper was written in 2022, we considered papers until 2021 in our search. In a first step, a combination of two keyword categories was applied to construct the search terms. To ensure an inclusive yet focused search, but also considering the healthcare domain the currently dominant application field of HDTs, we combined the search terms using (1) DT-related keywords, and (2) healthcare-related keywords (see Table \ref{Table1}).
\begin{table}[H]
 \caption{Search words used for identifying relevant literature in the context of Human Digital Twins. The keywords related to digital twins were combined with the keywords related to human factors and healthcare using the “AND”-connector.}
  \centering
  \begin{tabular}{p{0.5\textwidth} p{0.4\textwidth}}
    \toprule
    Keywords related to \textbf{digital twins}     & Keywords related to \textbf{human factors} and \textbf{healthcare} \\
    \midrule
    \multirow{3}{*}{Digital twin*} & Health* \\
                                & Medic* \\
                                & Patient* \\
    \multirow{3}{*}{Virtual twin*}  & Rehab* \\
                                & Human body* \\
                                & Human factor* \\
    \bottomrule
  \end{tabular}
  \label{Table1}
\end{table}

\begin{figure} [H]
  \centering
  \includegraphics[]{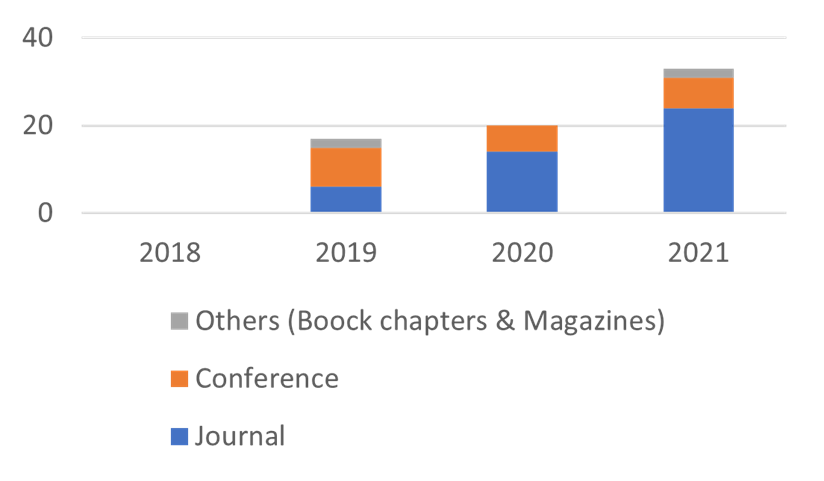}
  \caption{Distribution of publications relevant to the field of Human Digital Twins by year.}
  \label{Fig.4}
\end{figure}

Based on the resulting collection of papers, a first screening process was used to remove obviously out-of-scope papers. These include duplicates and obviously out-of-context papers based on title and abstract, and non-English papers. Further, only peer-reviewed conference and journal papers (or comparable) were considered, excluding books and not-retrievable publications. The remaining papers were then screened for their specific content to eliminate further out-of-scope papers whose content was not directly related to HDTs. For instance, many authors reported digital models and digital shadows as types of DTs, and were thus excluded due to their differences from DTs, as outlined in Section \ref{Section2.1}.\\
Finally, from the initial set of 2049 papers, a total of 70 research articles were considered relevant for further analysis. As illustrated in Figure \ref{Fig.4}, HDTs are still emerging as a major topic, with a growing number of papers per year. While in 2019, a total of 17 research papers (6 journal, 9 conference, and 2 book chapters) were published, this increased to 20 in 2020 (14 journal and 6 conference), and 33 in 2021 (24 journal, 7 conference, 1 book chapter, and 1 magazine article). Despite the increasing significance of HDTs in DT research, several review papers, such as \cite{barricelli2019, Fuller2020, Erol2020_1, Erol2020_2}, identified substantial unanswered research questions in this area.
Further, as illustrated in Figure 5, HDT research is currently found in four main domains: healthcare, industry, smart home and personal assistance, and sports. Here, healthcare applications are dominant, followed by the industrial domain. 
\begin{figure} [H]
  \centering
  \includegraphics[]{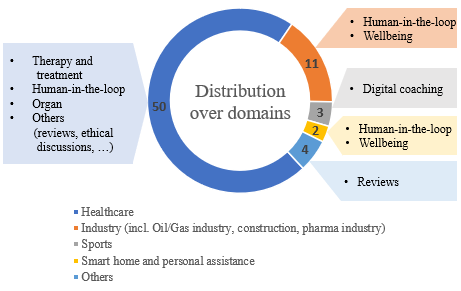}
  \caption{Distribution of publications relevant to Human Digital Twins across domains.}
  \label{Fig.5}
\end{figure}
In the healthcare domain, besides review papers (e.g., \cite{boulos2021, Mohapatra2020, wickramasinghe2021}), ethical discussions (e.g., \cite{braun2021, kerckhove2021, popa2021}), and other papers dealing with specific HDT applications, most papers describe the use of HDTs for therapy and treatment purposes. Here, HDTs are primarily used for the treatment of cancer (e.g., \cite{wickramasinghe2021, Schade2019, mourtzis2021}), diabetes (e.g., \cite{shamanna2020_1, Goodwin2020, shamanna2021}), multiple sclerosis (e.g., \cite{petrova2020, dillenseger2021, voigt2021}), post-spinal cord injury (e.g., \cite{pizzolato2019, pizzolato2021}), and in gait therapy (e.g., \cite{dillenseger2021, voigt2021}). Additionally, some healthcare papers describe HDTs as human-in-the-loop DTs (see the following section), where the human is considered part of a superordinate healthcare system (e.g., \cite{nonnemann2019, rodriguez-aguilar2020, firouzi2021}), or break down HDTs as representations of specific organs (e.g., \cite{Lauzeral2019, martinez2019, corral2020}).\\
In the industrial domain, HDTs mainly represent human workers within specific working processes, such as Human-Robot-Collaborative tasks (e.g., \cite{Savur2019, lv2021, wangT2021}). Further, several papers suggest initial approaches to using HDTs for improving employees’ well-being (e.g., \cite{wangZ2021, wanasinghe2021, montini2021}).\\
In the sports domain, HDTs are solely used for virtual coaching purposes (e. g. \cite{diaz2020, barricelli2020, diaz2021}), whereas in the Smart home and personal assistance domain, HDTs are implemented as virtual assistants to improve users’ general well-being (e.g., \cite{abeydeera2019}) and align users’ needs with other smart devices (e.g., \cite{Hafez2019}).\\
As illustrated in Figure \ref{Fig.5}, even within the HDT literature, several different HDT applications exist, each with different requirements and challenges for HDT design, demanding a closer examination of the fundamental elements of HDTs. Thus, to address our research questions stated in Section \ref{Section1}, we reviewed the included papers for a common understanding of a cross-domain definition, key characteristics, and emerging challenges of HDTs, that arise from applying the concept of DTs to a human entity across various applications.

\section{Human Digital Twin - Definitions, Categorization, and Emerging Challenges}
\label{Section4}
Current literature does not provide a consistent definition of HDTs. In addition, various authors describe HDTs that differ substantially from the fundamental definition of DTs or correspond to related concepts, such as the digital model or digital shadow (see Figure \ref{Fig.1}). Therefore, to create a common understanding of HDTs, the following sections distill general characteristics and emerging challenges from the reviewed cross-domain literature. At this point, it should be noted that the following results have been aggregated from across papers through comparison and contrast, as well as assessment of the general context and discussion, without individual papers necessarily explicitly stating results. Table \ref{Table4} (Appendix) provides a detailed overview of the reviewed papers in terms of their application domain, year of publication, publication type (C = Conference, J = Journal, B = Book chapter, M = Magazine), content, and their contribution to the challenges ("\checkmark" indicates a strong focus or direct addressing whereas “(\checkmark)” indicates indirect derivability).

\subsection{Definition and Categorization}
\label{Section4.1}
\begin{figure} [H]
  \centering
  \includegraphics[]{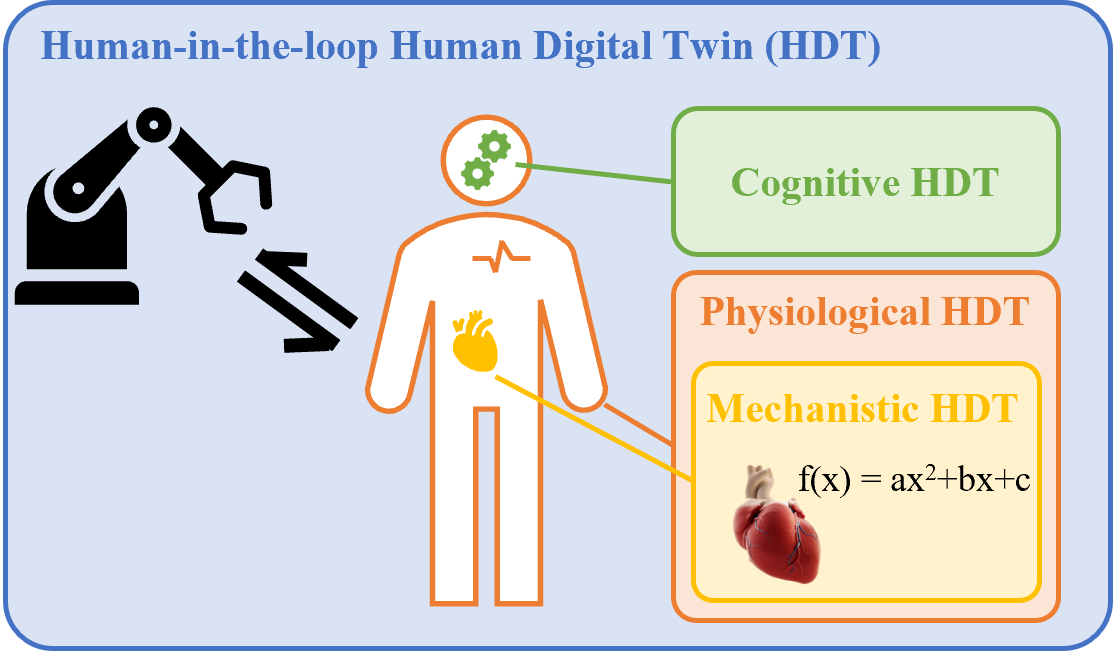}
  \caption{The four categories of Human Digital Twins and their correlation.}
  \label{Fig.6}
\end{figure}

In 2018, the definition of DTs was extended to include the replication of living or nonliving physical entities \cite{diaz2020}. This paved the way for the development of HDTs, which in literature—depending on the respective application—are also referred to as Personal DT (e.g., \cite{Bagaria2020}) or DT for Healthcare (e.g., \cite{Angulo2020}), among others. Despite the growing number of research entries about HDTs, there are currently few general, cross-domain definitions, with inconsistent use of the concept within different applications. For instance, \cite{kerckhove2021} describes HDTs as “[...] a twinning process that addresses the whole person and his or her history and biological condition.” However, the generality of such definitions limits their informative value. In contrast, more concrete definitions are limited to a specific use case and therefore do not directly contribute to a general understanding of the concept behind HDTs. \cite{rivera2019}, for example, defines HDTs as “[...] a set of interrelated and virtual anatomical structures (AS), which might represent tissues, organs, or systems in the human body”, whereas \cite{Bagaria2020} describes them as a “[...] data-driven technology that reflects the health status of individuals inferred from the continuously collected data.” These definitions are only valid in a medical context and do not necessarily apply to HDTs in other domains. Thus, there is a need to identify essential characteristics of HDTs across application domains and fuse them into a new general yet comprehensive cross-domain definition.\\
As a basis for distilling such characteristics, we identified four principal categories of HDTs in the literature (see Figure \ref{Fig.6}).\\

First, \textbf{human-in-the-loop HDTs} represent humans as part of superordinate systems, such as medical care units or human-machine collaboration systems. As such, the human’s state is typically defined by their location, pose, or general health state. For instance, \cite{shaked2021} proposes a hospital management system representing patients alongside other interconnected functional entities by their health status (body temperature) and location to support decision-making. In contrast, \cite{lv2021} and \cite{tzavara2021} combine RGBD cameras, human pose estimation tools (e.g., OpenPose \cite{openpose}), and deep learning to detect the human’s body posture, the intended movements, and correspondingly control the robotic system for a safe and efficient human-robot collaboration. Therefore, human-in-the-loop HDTs focus primarily on the human’s integration into the superordinate system, thus representing the human only through very particular aspects directly relevant to the system.\\
Second, \textbf{physiological HDTs} represent the human’s physical condition based on biomarkers. The selection of physiological aspects represented by the HDT thereby varies depending on the corresponding use case. \cite{derungs2020}, for example, describes the use of inertial sensors to analyze movements in gait rehabilitation, whereas \cite{pizzolato2019} proposes a whole musculoskeletal patient model for spinal cord injury rehabilitation. In contrast, other use cases might require vital biomarkers instead of musculoskeletal factors. \cite{shamanna2020_1} and \cite{shamanna2021}, for instance, describe how HDTs that monitor blood pressure and blood glucose level can be used in diabetes treatment, whereas other vital parameters, such as the heart rate, can be relevant for HDTs in virtual coaching applications (e.g., \cite{barricelli2020}) or to monitor the well-being of employees in industry (e.g., \cite{Savur2019}). Thus, the human’s physiological condition is multifaceted and can include a variety of factors from musculoskeletal to vital functions.\\
Third, \textbf{mechanistic HDTs} represent the physiological aspects of the human through mechanistic models. Especially on the organ and cellular level, sensors often cannot sufficiently perceive certain aspects of the human body. However, in many cases, the sensor data can be used to continuously parameterize models to describe these functionalities instead. While \cite{corral2020} describes how Navier Stokes equations, for example, can be used to represent blood flow, \cite{Goodwin2020} point out the use of differential equations to describe the blood glucose level response to particular drugs. Further, \cite{Lauzeral2019} proposes data-based Parametric Reduced Order models that generate real-time adaptive models of organs. Thus, mechanistic HDTs include digital models that are continuously updated with indirect physiological sensor data to represent specific aspects of the human body, especially on the organ or cellular level.\\
Finally, \textbf{cognitive HDTs} represent the human’s cognitive mechanisms. The range of cognitive properties that can be incorporated into a cognitive HDT is diverse. \cite{petrova2020}, for example, proposes an HDT for multiple sclerosis therapy that incorporates the patient’s processing capabilities, memory, expressive language, and executive functions. Since many of these cognitive mechanisms cannot yet be assessed automatically using sensors, qualitative evaluation via standardized manual tests is often still required \cite{petrova2020}. Other cognitive aspects, such as stress or mental fatigue, can be derived from, for example, physiological sensor data such as the heart rate. Thus, in contrast to the other categories (Figure \ref{Fig.6}), cognitive HDTs focus on non-physical aspects that often cannot be measured via sensors, and therefore represent one of the most significant differences from traditional DTs.\\
Although there are basic commonalities, these four categories highlight several fundamental differences between HDTs and traditional DTs. Most notably, although HDTs—similar to traditional DTs—rely on continuous synchronization with the physical entity, the human body does not contain integrated sensors, making it often impossible to perceive certain aspects directly. In addition, HDTs add entirely new aspects to the concept of DTs, such as cognitive mechanisms. These differences create a number of distinctive challenges for HDTs.

\subsection{Emerging Challenges}
\label{Section4.2}
HDTs differ from traditional DTs in several important ways, resulting in new technical and ethical challenges. Although some HDT literature already points directly to such challenges, comparing papers across domains reveals a number of additions (e.g., through conflicting statements about HDTs and their use). Bringing these findings together, we identify three main areas of emerging challenges.

\subsubsection{Technical Conceptualization}
\label{Section4.2.1}
Generally, HDTs follow the same technological principles as traditional DTs. Thus, HDTs typically include sensor technology to perceive the human entity, process and store the data, and provide feedback to the human user. However, they also have several significant differences from traditional DTs, partly driven by technical challenges due to human factors and the novelty of the concept itself.\\
First, human entities are highly \textbf{complex}. In contrast to traditional DTs, such as product DTs, which often represent systems limited in the number of parameters and attributes, the human body is a significantly more complex physical system \cite{boulos2021}, which introduces cognitive (e.g., \cite{petrova2020}), contextual (e.g., \cite{wangT2021}), and social elements (e.g., \cite{Comito2019}). \cite{Schade2019}, for example, emphasizes that cancer treatment applications require the HDT to represent the human on a cellular, thus highly detailed and hardly perceivable level. In contrast, \cite{petrova2020} focuses on monitoring the human’s cognitive capabilities, which again consist of various highly-complex mechanisms, such as memory and language. Apart from such user-internal aspects, \cite{martinez2019} further emphasizes the importance of additionally considering contextual factors, such as the user’s social behavior, e.g., by exploiting social media data. Thus, humans pose a highly complex system with multiple abstraction levels, leading to a natural limitation regarding the possible fidelity of its digital representations.\\
Second, human entities are highly \textbf{dynamic}. Humans themselves (and all of their complex sub-elements), as well as the environmental entities affiliated with them, can change at any time. For example, in healthcare, \cite{dillenseger2021} emphasizes the rapid changeability of patients’ health states, and \cite{rodriguez-aguilar2020} and \cite{Calderita2020} emphasize the dynamic interaction between patients and various entities in their environment, such as doctors, caregivers, equipment, and other environmental objects. Further, \cite{Hafez2019} describes the deep connection between HDTs and their surrounding smart devices in a smart-home environment. In industry, HDTs are often in collaboration with DTs of equipment \cite{lv2021}. Thus, changes within these connected devices or humans themselves significantly challenge the flexibility of HDTs and their design.\\
Third, the action of human entities depends on \textbf{trust and motivation}. \cite{maeyer2020} shows that many users associate HDTs with a collection of data linked to specific algorithms. In fact, many HDT applications align with this understanding. \cite{montini2021} and \cite{Calderita2020}, for example, conceptualize their HDT as user profiles that only run in the background and provide data to prediction tools and other system components. Although such conceptualization might be sufficient for some applications, skepticism exists about their suitability for applications that depend on human users acting based on feedback from the HDT, which requires trust and motivation. For example, in elderly care or interactive healthcare, these are particularly critical aspects of HDT performance \cite{maeyer2020}. Thus, the HDT’s effectiveness highly depends on its perception by the user, building the foundation for trust and acceptance.\\
Fourth, human entities place a high degree of \textbf{constraint} on technologies that might help resolve some of these complexity issues. For example, \cite{popa2021} and \cite{firouzi2021} highlight the limits in perceiving particularly internal aspects of the human body given current sensor technology. Further, due to the intrusiveness of external sensors, a permanent data connection between the physical entity and the HDT in most cases is not feasible, leading to inconsistencies in the available data \cite{popa2021, barricelli2020}. Although mechanistic or statistical models can compensate for lacking real-time data by predicting the human body’s behavior, \cite{corral2020} emphasizes those models’ constraints by the assumptions and principles they build on, and the underlying base of data. Thus, the combination of complexity, dynamism, and technological constraint places significant limits on the fidelity of HDTs.\\
Finally, more practically, there is currently little \textbf{standardization} across HDT applications. In addition to the direct interaction with their human entity, the connection of HDTs with other DTs allows for enhanced interaction with their environment, for example, through alignment and knowledge exchange. In industry, \cite{Savur2019} and \cite{tzavara2021}, for example, suggest connecting HDTs with robots’ DTs to enable safe human-robot collaboration by aligning the human’s intentions with the robot’s movements. In healthcare, interlinking patient HDTs of similar medical conditions can promote more robust decision support and faster adaptability through knowledge exchange \cite{dillenseger2021}. To enable such communication, however, HDTs need to use standardized data communication methods for seamless data transmission \cite{elayan2021}. Despite the increasing number of DT applications, only a few approaches pursue a standard-oriented implementation, significantly complicating the interconnection of HDTs and their environment.\\
Together, these five challenges serve to highlight the specific differences between the technical conceptualization of HDTs and DTs.

\subsubsection{Perception and Acceptance}
\label{Section4.2.2}
While all DTs should be trusted by their users, HDTs place significant additional demands on perception and acceptance. In particular, transparency and trust play a distinctive role in HDT applications. Not only can a lack of trust hinder the adoption process itself \cite{chase2021}[76], but it can also influence the user’s engagement with the system. In many applications, such as therapy and rehabilitation, a lack of engagement has a significant impact on the outcomes \cite{voigt2021}. In HDT applications, transparency and trust challenges can be understood in relation to the three layers of the HDT \cite{elayan2021}.\\
First, in the data collection layer, low \textbf{data quality and consistency} pose a significant challenge. High-quality data are important for reliable decision-making. However, due to the dependency on external and—often intrusive—sensors, HDTs usually don’t perceive continuous data, resulting in data gaps and inconsistencies \cite{popa2021, barricelli2020}. Further, \cite{petrova2020} reports difficulties in the automatic assessment of cognitive mechanisms, resulting in a dependency on manual data. \cite{Laamarti2020}, however, emphasizes the low reliability of manual data due to human errors and other factors. Thus, low-quality and inconsistent data can severely affect the system’s performance and potentially endanger the user’s trust in the HDT system.\\
Second, in the data storage and processing layer, apart from storage capacity and security challenges known from traditional DT and big data applications, \textbf{transparency} in decision processes is the main challenge. Mechanistic HDTs, for example, often consist of complicated equations that lack interpretability \cite{corral2020}. Similarly, \cite{barricelli2020} highlights the lack of transparency of AI-based HDTs, given the “black box” nature of AI. Further, \cite{Fuller2020} and \cite{chase2021} point out additional skepticism derived from the prevailing fear that AI could surpass humanity and undermine or replace professionals in the future. Thus, insufficient transparency within the storage and processing layer can introduce trust issues in the interaction between the HDT and the user.\\
Third, in the output and interaction layer, \textbf{comprehensibility} of the presented information can lead to challenges of confusion and frustration. Here, various authors (e.g., \cite{barricelli2019, Fuller2020, Laamarti2020}) highlight the significance of visualizing HDT data intuitively and understandably. However, intuitive visualization can be challenging with increased data complexity \cite{barricelli2019}. \cite{shaked2021}, for example, illustrates how reasonable grouping and structuring of visual elements require careful consideration, especially on systems consisting of multiple entities. Further, the user interface proposed in \cite{nonnemann2019} highlights the necessity to distribute information over multiple menus to maintain an overview. Thus, the user interface has the potential to promote or undermine the transparency and understandability of the HDT system.\\
Finally, \textbf{regulatory compliance} is a critical cross-layer challenge, which can impact trust. Regulations concerning HDTs typically serve user protection purposes. For HDTs in healthcare applications, \cite{Lutze2020}, for example, points out the EU Unique Device Identification system guidelines, which require the assignment of unique identification numbers to provide traceability of medical products. Further, \cite{corral2020} mentions the General Data Protection Regulation (GDPR), which ensures users full control over the use of their data. Such regulations can constrain the development process, but can also raise significant trust issues if not considered.\\
Together, these four challenges serve to highlight the specific trust-related aspects, particularly impacting the perception and acceptance of HDTs in contrast to traditional DTs.

\subsubsection{Ethics}
\label{Section4.2.3}
Many challenges regarding transparency and trust result from the concern for user-related consequences. This makes ethical issues uniquely important in HDT design.\\
First, \textbf{data sensitivity} is a major challenge in any application working with human data, especially in the medical domain. Data theft or misuse can lead to severe consequences for the users \cite{barricelli2019, diaz2020}. \cite{maeyer2020}, for example, reveals serious concerns of users about their personal data’s security, highlighting that most users would only want to share data with medical staff and stakeholders directly relevant to the therapy. In addition, when stored safely, the user data could still be misused by the parties involved. \cite{kenzierskyj2019} highlights that insurance companies, for example, could use patients’ health data to adjust their service or cost structures to their disadvantage. Further, besides the security and use of user data throughout the HDT’s application, there are questions about what to do with the HDT and its underlying data in the event of an (unexpected) termination of the application. In elderly care, \cite{maeyer2020}, for example, indicates that some users wish their HDT to be deleted in the event of their death, whereas others could imagine making their data available to contribute to the further success of the system. Thus, despite the indisputable significance of data security, opinion on data ownership varies, requiring a high degree of sensitivity and flexibility regarding the users’ demands.\\
Second, in addition to the security of personal data, also \textbf{user safety} itself plays a critical role in the application of human-centric technologies of all kinds. Particularly HDTs have a significant impact on the user's physical integrity due to their physical feedback and the decisions made based on their support. \cite{pizzolato2019}, for example, warns that incorrect strength and timing of feedback applied through powered exoskeletons can lead to pain or even injury. Further, \cite{Erol2020_2} highlights that mistakes in the detection of user-related findings or their misinterpretation can lead to incorrect diagnoses and, correspondingly, to wrong treatment. Thus, HDTs have the potential to pose a significant risk to user's well-being due to their direct or indirect interaction with the user.\\
Third, the potential \textbf{societal impact} of HDTs is a sensitive issue. HDTs can promote (or hinder) social equality, particularly concerning equal access to knowledge and services. In e-commerce, for example, HDTs could suggest optimal products matching the user’s preferences, promoting independence of financially or politically influenced suggestion systems \cite{Erol2020_1}. \cite{popa2021} further hints at the opportunity to outsource part of medical therapies from the clinical to the patient’s domestic environment through telemedicine, reducing the patient’s dependence on the availability of clinicians and, conversely, relieving the clinical professionals. Challenging these opportunities, \cite{dillenseger2021} indicates that some social groups—or even countries—may not be able to afford the necessary equipment or know-how needed for such HDT applications. Thus, HDTs have the potential to compensate for or reinforce social differences, depending on their availability and equality in distribution and use.\\
Fourth, apart from equity in availability, equal \textbf{representation} amongst potential user groups is also critical. Here, several authors hint at the risk of bias and discrimination in case of insufficient representation of particular user groups. \cite{corral2020}, for example, argues that unbalanced data could lead to racial or social under- or misrepresentation. \cite{boulos2021} goes one step further, indicating that HDTs could identify and represent patterns across social groups and thus enforce discrimination. \cite{shamanna2020_1} also points out the potential for sexism, arguing that using a seductive female voice in the course of verbal interaction could contribute to the reinforcement of sexist stereotypes. Thus, HDTs can potentially promote critical social challenges.\\
Finally, several authors highlight the risk of \textbf{dependency}. HDTs show great potential in supporting users, for example, through decision support and the possibility of self-monitoring. However, these benefits also bring certain risks. \cite{kerckhove2021} and \cite{rodriguez-aguilar2020}, for example, warn that outsourcing knowledge and decisions to HDTs could result in a loss of the user’s competencies caused by the lack of practice. Apart from this, \cite{boer2020} argues that the enhanced focus on particular aspects of the human being could also lead to neglecting other essential factors not represented by the HDT. Thus, the relationship between the HDT and its involved stakeholders is a critical source of ethical issues.\\
Together, these five challenges emphasize the unique ethical issues raised by HDTs, which go far beyond questions of technical feasibility, acceptance, and the benefits of the individual user.

\section{Discussion and Implications}
\label{Section5}
The results in Section \ref{Section4} provide the foundation for answering our research questions. Thus, this section i) formulates a general, cross-domain definition of HDTs based on the identified characteristics and categories, ii) introduces key considerations for the design of HDTs that emerge from the distilled challenges, and iii) discusses these findings’ limitations. Finally, we translate the generated insights into a general HDT framework and discuss their main practical and theoretical implications as well as the resulting need for future research.

\subsection{Definition of Human Digital Twins}
\label{Section5.1}
\begin{figure} [H]
  \centering
  \includegraphics[scale=0.8]{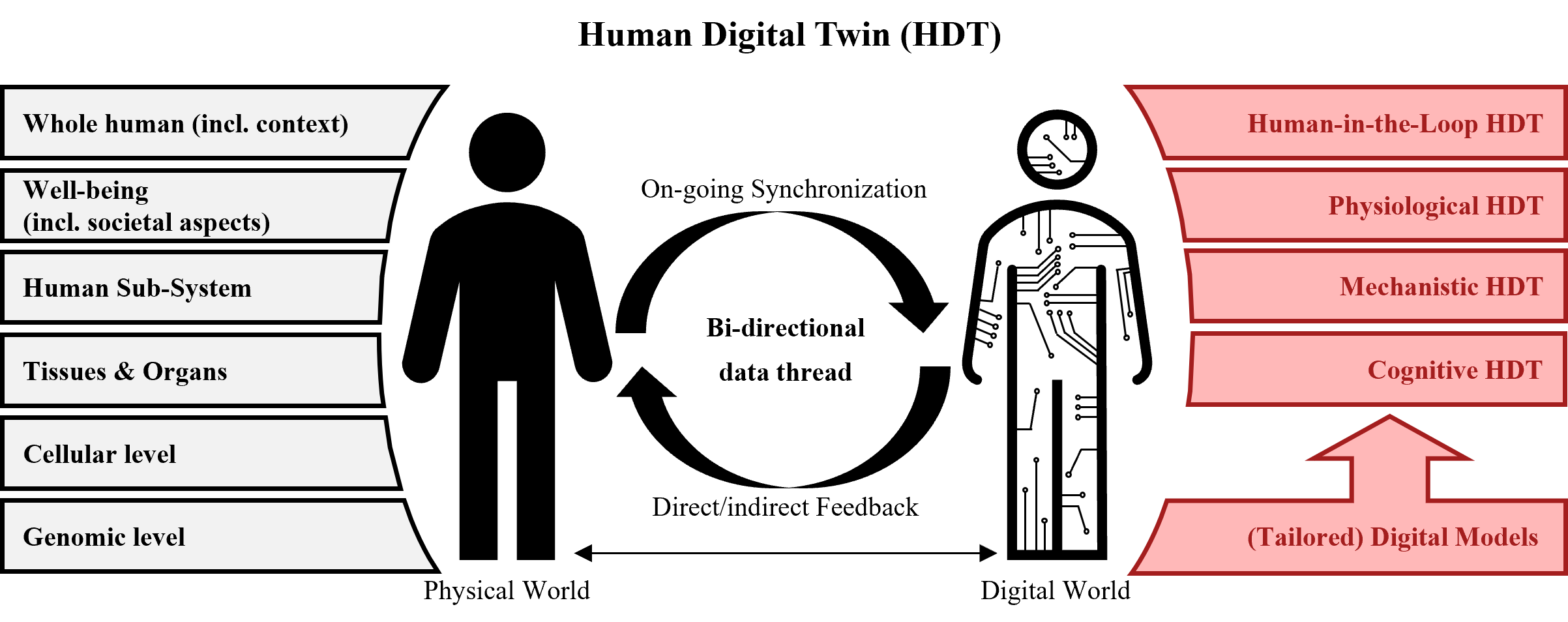}
  \caption{A general conceptualization of Human Digital Twins defined as: A digital representation of a human (or specific aspect of a human); which uses a bi-directional data thread for on-going synchronization with the human’s state; and provision of feedback based on data aggregation and predictions to directly (or indirectly) influence the human entity, user(s) and/or their environment.}
  \label{Fig.7}
\end{figure}

The results of the review reveal both the absence of a clear definition of HDTs, and common features. Thus, this section merges insights from across papers to formulate a first general, cross-domain definition of HDTs.\\
First, the categories of HDTs identified in Section \ref{Section4.1} show a division of the human body into levels of abstraction. This raises the question as to which abstraction level to consider a DT representing a specific human aspect an HDT. \cite{kerckhove2021} argues that systems that gather too specialized and fragmented data cannot sufficiently represent the user to enable relevant data analytics. However, \cite{braun2021} and \cite{popa2021} counter-argue that the required fidelity of the HDT should be context-specific and thus determined by the specific use case. Further, \cite{Bagaria2020} states that HDTs can represent humans at various abstraction levels ranging from single-body parameters to the human as a whole. To date, developing holistic HDTs that include all human aspects in full detail—given their high complexity—is impossible \cite{Goodwin2020}. Therefore, focusing on a specific aspect of the human entity, such as specific body parts, organs, or other physiological or cognitive attributes, is essential. Thus, HDTs should include all DTs representing a use case-specific aspect of the human entity independent from its abstraction level.\\
Second, the traditional definition of DTs requires continuous synchronization of the HDT with its physical entity, typically through a permanent bi-directional data thread (see Section \ref{Section2.1}). However, as indicated in Section \ref{Section4.2.1}, permanent synchronization in most HDT cases is impossible due to the absence of integrated sensors. Therefore, several papers excluded from this review (e.g., \cite{greenbaum2021}) argue that HDTs do not necessarily need to synchronize with the human’s state but that it is sufficient for them to predict it reliably. In contrast, \cite{braun2021} and \cite{canzoneri2021} clarify that simulating the human’s state is not enough to call a system an HDT since these must be able to adapt to changes in the human’s condition. However, it is debatable as to what extent the synchronization must be continuous. While specific use cases entail rapid changes in the human’s condition and thus require frequent synchronization (e.g., \cite{dillenseger2021}), others only require sporadic updating of the HDT (e.g., \cite{abeydeera2019}). Either way, the synchronization must be on-going either with regular or irregular intervals, as a one-time parameterization would tailor the HDT to the human entity but not provide adaptability to changes in the human’s or environmental condition. Therefore, HDTs must provide on-going synchronization with a frequency determined by the specific use case.\\
Third, a fundamental characteristic of DTs is the ability to provide feedback to the physical entity. Across the reviewed papers, HDTs provide various types of feedback. However, in contrast to the autonomous control feedback that DTs provide, HDTs often do not directly control the human user but instead influence them indirectly, for example, by providing decision support (e.g., \cite{barricelli2020}). This raises the question of what the minimum requirement for the feedback provided by HDTs is. According to the definitions stated in Section \ref{Section2.1}, the key difference between DTs and digital shadows lies in any kind of feedback that makes the DT go beyond “just monitoring” the physical entity. Therefore, HDTs should provide any kind of haptic, acoustic, visual (or other) feedback, that is based on aggregated data and directly or indirectly influences the human entity or its environment.\\
Bringing these insights together, we are able to distill a general definition of HDTs across all current application domains:\\
\\
\textit{A digital representation of a human (or specific aspect of a human); which uses a bi-directional data thread for on-going synchronization with the human’s state; and provision of feedback based on data aggregation and predictions to directly (or indirectly) influence the human entity, user(s) and/or their environment.}\\
\\
In addition to this definition and its underlying characteristics, HDTs are further characterized with respect to four major application categories. Here, HDTs represent human entities within superordinate DT systems (human-in-the-loop HDTs), physiological attributes (physiological HDTs), physiological attributes represented through mechanistic models (mechanistic HDTs), and/or cognitive attributes (cognitive HDTs) (see Figure \ref{Fig.6}). Although these categories represent all aspects of the human entity, they do not always strictly differentiate from each other. For instance, mechanistic HDTs often also represent physiological aspects of the human entity, thus extending physiological HDTs. Further, HDTs are not limited to a single category. A single HDT could also represent various human aspects, thus combining multiple HDT categories, resulting in “composite HDTs” \cite{boulos2021}. In behavior-changing therapy, for example, an HDT would have to represent both the patient’s physical as well as cognitive capabilities and impairments to provide decision support concerning all relevant aspects. Therefore, these categories complement our strict definition by providing a descriptive framework for understanding the general layers of HDT applications, as summarized in Figure \ref{Fig.7}.

\subsection{Challenges of Human Digital Twins}
\label{Section5.2}
Based on the understanding outlined in Figure 7, it is possible to identify 14 main challenges emerging from the literature, as summarized in Table \ref{Table2}.\\

{\centering
\begin{longtable}{p{0.3\textwidth} p{0.6\textwidth}}
 \caption{14 key challenges in Human Digital Twin (HDT) design.}
 \label{Table2}\\
    \toprule
    Challenge     & Description \\
    \midrule
    \multicolumn{2}{l}{\textit{Technical conceptualization}} \\
    \midrule
    Complexity & Humans are highly complex physical systems that besides consisting of multiple possible abstraction levels include cognitive, contextual, and social elements. Thus, a natural limitation exists regarding the possible fidelity of its digital representations.\\
    Dynamics  & Humans and their environments are highly dynamic systems where the human’s condition, connected devices, and the application’s context can change frequently. Thus, changes within the human itself or within its environment significantly challenge the flexibility of HDTs and their design. \\
    Trust and motivation  & The effectiveness of HDTs, particularly in applications with direct user interaction, highly depends on their perception by the user. Thus, the HDT’s conceptualization can support or undermine the user’s trust and motivation. \\
    Technological constraint & Humans cannot be permanently measured in full detail due to the lack of integrated sensors and humans’ sensitiveness concerning intrusive technologies. Thus, such technological constraints place significant limits on the fidelity of HDTs.\\
    Standardization & HDT applications often require communication between the HDT and other systems, such as sensors or DTs of its environment. Thus, the lack of standards within HDT design and communication protocols significantly complicates the interconnection of HDTs and their environment.\\
    \midrule
    \multicolumn{2}{l}{\textit{Perception and Acceptance}} \\
    \midrule
    Data quality and consistency & The reliability of the HDT’s decision-making relies on the quality and consistency of the underlying data. Thus, low-quality and inconsistent data resulting from the HDT’s dependency on external and – often intrusive – sensors impacts the system’s performance and potentially endangers the user’s trust in the HDT system.\\
    Transparency  & HDTs often rely on complex equations or black box AI systems which limit the understandability of the resulting decision-making. Thus, insufficient transparency within the HDT can introduce trust issues in the interaction between the HDT and the user. \\
    Comprehensibility  & HDTs represent large amounts of complex data that lead to confusion and frustration if not appropriately represented. Thus, a poorly designed user interface limits the transparency of the HDT system and could impact its comprehensibility and trust by the user. \\
    Regulatory compliance & HDTs underlying regulations can constrain the development process but are eminent in terms of user protection. Thus, not complying with existing regulation can lead to problems with authorities and raise significant trust issues, yet regulation in this context is still formative due to the emerging nature of the technology and research.\\
    \midrule
    \multicolumn{2}{l}{\textit{Ethics}} \\
    \midrule
    Data sensitivity & HDT-related data are particularly sensitive as its misuse or theft can result in direct damage to the human user. Thus, HDTs require a high degree of sensitivity and flexibility regarding the users’ demands.\\
    User safety  & Incorrect feedback can lead to wrong decisions or even cause physical harm to the user. Thus, HDTs pose a potential risk to the user's well-being due to their direct or indirect interaction with the user.  \\
    Societal impact  & HDTs require expensive equipment, advanced IT infrastructure, and expert knowledge, which might not be affordable or available for specific user groups or countries. Thus, HDTs have the potential to compensate or reinforce social differences, depending on their availability and equality in distribution and use.\\
    Representation & HDTs are often based on limited data that might not represent all user groups equally. Thus, HDTs can potentially promote critical social challenges such as bias or discrimination, depending on the underlying data.\\
    Dependency  & HDTs raise awareness and provide decision support for specific human aspects, which can lead to the neglect of other important aspects and the outsourcing of decision-making capabilities from the user to the HDT. Thus, HDTs can promote dependency and loss of capabilities through lacking practice of the user.  \\
    \bottomrule
\end{longtable}}
Despite some of these challenges, such as high complexity, dynamics, and data sensitivity, being widely acknowledged (e.g., \cite{boulos2021,rivera2019,Angulo2020}), the majority have not been widely discussed but only appear in specific papers or arising from the analysis of applications. For example, we highlight key challenges in perception, trust, and ethics. Further, some challenges appear similar to challenges known from traditional DTs. However, the differences between HDTs and traditional DTs significantly affect how the challenges impact design. Challenges in HDT design are rooted in three key differences from traditional DTs.\\
First, HDTs position humans as the observed physical entity instead of human-made products, processes, or systems. This raises entirely new challenges connected to human factors. Particularly the human’s complexity, sensitivity, perception, and trust significantly differentiate HDTs from traditional DTs.\\
Second, unlike most traditional DTs, HDTs mainly rely on external sensors and manual inputs instead of integrated sensors. As outlined in Table \ref{Table2}, these sensors can be intrusive and often do not fully observe the human entity, resulting in decreased data availability and consistency in HDT applications compared to traditional DTs.\\
Third, the novelty of the HDT concept results in many important factors not being explored yet by research. This manifests in a noticeable lack of common practices, understandings, and established technologies.\\
Therefore, the differences between HDTs and traditional DTs lead to a number of specific and unanswered challenges that demand further guidance, for example, in clarifying their impact on the design of HDTs.

\subsection{Design Considerations}
\label{Section5.3}
The previous section discussed the main challenges arising from the HDT concept and the emerging demand to clarify their impact on HDT design. Thus, this section derives key design considerations, suggests possible implementations based on the reviewed HDT literature, and discusses their effect on the application and the user.

\subsubsection{Data Collection and Processing}
\label{Section5.3.1}
Three key design considerations emerge regarding the collection and processing of data within the HDTs.\\
First, HDTs must represent the human entity with sufficient \textbf{fidelity and accuracy}. These attributes concern the level of abstraction and the accuracy of representation, which both depend on the use case-specific requirements. One of the most critical challenges for accurately representing the human entity is that current sensor technology cannot perceive all aspects of the human body directly. However, \cite{braun2021}, amongst others, suggests combining (mechanistic) models that describe the observed human aspect with sensor data measuring other human aspects correlated to it. This way, the HDT can derive information about the observed human aspect and correspondingly reparametrize the underlying model. Thus, the combination of modeling and sensor data is one option to address complexity-related challenges and provide sufficient fidelity and accuracy within the HDT.\\
Second, HDTs must provide \textbf{sensitivity and robustness} concerning inconsistent data. To still provide high accuracy and reliability, HDTs must compensate for data inconsistency by filling data gaps and increasing the overall data quality. \cite{petrova2020} and \cite{barricelli2020} highlight the possibility of augmenting missing data by either adopting them from similar data sets or generating them from their similarity to previously perceived data. Alternatively, \cite{patel2017} emphasizes the possibility of calculating missing data using corresponding formulas. Another recent method for data augmentation stated in the broader literature in data analysis is generating synthetic data, for example, through Generative Adversarial Networks \cite{poudevigne-durance2022}. Despite their promising capabilities to augment data reliably in most cases, the mentioned methods depend on the availability of high-quality data themselves. In addition, traditional data processing methods such as filtering can be used to increase the quality of existing data \cite{alcaraz2019}. Thus, data processing capabilities can compensate for inconsistent and low-quality data, improving the overall reliability, and thus the trustworthiness of HDTs.\\
Together, corresponding models and pre-processed sensor data fusion have the potential to allow HDTs to address complexity-related challenges and accurately represent even non-perceivable human aspects.

\subsubsection{Flexibility and Interconnectivity}
\label{Section5.3.2}
Two key design considerations emerge regarding the flexibility and interconnectivity of HDTs.\\
First, HDTs must promote \textbf{interconnectivity} with other DTs. Interaction and data exchange with other DTs can provide the HDT with additional data, thus increasing its reliability and robustness. For example, interlinking patient HDTs could lead to sharing of experiences and thus more robust diagnostics \cite{yadav2018}. However, there are significant challenges in communication between DTs, mainly due to varying data formats. One way to tackle these challenges is to use common system behavior modeling tools such as SysML (e.g., \cite{shaked2021}) or ROS (e.g., \cite{tzavara2021}) that allow modeling all DTs of one DT network as well as their interactivity within the same modeling environment. However, this approach has limitations in applications that include DTs from various developers. Alternatively, HDTs can use primarily standardized communication protocols. \cite{martinez2019}, for example, suggests using the IEEE X73 standard, which promotes standardized communication protocols for medical devices. This way, received data do not need to be converted but can be used directly, improving the HDTs simplicity and efficiency. Thus, using standardized communication protocols allows for seamless interaction between DTs, extending the HDT’s environmental awareness and thus their accuracy and robustness.\\
Second, HDTs must provide \textbf{adaptability} regarding changes within the human entity and its environment. HDTs need to detect these changes and adapt to them accordingly. Therefore, \cite{barricelli2019}, for example, suggests using evolving AI capable of continuously adapting the HDTs decision-making to new data. Further, the broader literature highlights modularity as a key design feature of DTs, enabling the developers to adapt the DT to changes that require fundamental structural modifications, for instance, through new regulations. \cite{masison2021}, for example, suggests dividing DTs into modules that only communicate through a central data space. This way, modules can dynamically be added or deleted without considering dependencies. Thus, while self-learning algorithms can adapt the HDTs decision-making to new data, a modular design can provide adaptability to more fundamental changes.\\
Together, increased data availability through interconnectivity to other DTs, adaptability through continuous learning techniques, and modular design have the potential to significantly improve the HDTs fidelity and accuracy, thus allowing for the representation of even complex dynamic relationships.

\subsubsection{Transparency and Trust}
\label{Section5.3.3}
Four key design considerations emerge regarding the transparency and trust of HDTs.\\
First, HDTs need to provide \textbf{identifiability and traceability}. Since the feedback of HDTs relies on underlying user data, confusing HDTs would result in wrong feedback for the user. Therefore, besides being required by certain regulations, such as the EU Unique Device Identification system guidelines, assigning each HDT a unique identification number is one way to prevent such confusion \cite{lutze2019}. In addition, the aforementioned dynamics in HDT applications might require significant changes within the HDT. \cite{lutze2019} therefore calls for including versioning to track fundamental changes, such as security patches. Thus, HDTs that contain unique identifiers and version numbers reduce the risk of confusion and promote transparency through the traceability of the HDT’s evolution.\\
Second, \textbf{credibility and explainability} of HDTs and their feedback are crucial for the user’s trust. Due to the high complexity of HDTs and the resulting challenges concerning their understandability by non-expert users, the feedback provided by the HDT needs to be explainable and justifiable. To face this challenge, the majority of papers, for example, \cite{diaz2020} and \cite{dang2021}, call for the integration of experts and professionals in the HDT’s design and application, as well as the incorporation of profound domain knowledge, for example, in the development of included AI systems. However, the resulting dependence on such domain knowledge and experts could lead to new trust barriers, depending on their credibility \cite{popa2021}. \cite{diaz2020} and \cite{dang2021} further suggest extensive testing of the HDT system on its accuracy and reliability. Using profound validation metrics could prove the HDTs proper functionality and allow comparison with related systems. Therefore, while integrating professionals and domain knowledge helps to face complexity challenges in the development process, testing the resulting system using profound validation metrics proves the HDT’s reliability and thus improves its credibility.\\
Third, the \textbf{embodiment} of HDTs is a new conceptualization approach that could improve the HDT’s graspability. Within the broader information system literature, several authors highlight the potential of embodying complex systems into graphical or physical entities to improve the system’s perception by its user(s) \cite{aljaroodi2019}. Therefore, especially in applications with direct HDT-user interaction, integrating HDTs into visual or other easily graspable entities could positively affect the HDT’s transparency and understandability, and could even allow for bond-forming between the user and their HDT.\\
Most notable in this context is the integration of HDTs into human-like user avatars. Embodying HDTs into avatars extends the traditional HDT conceptualization from only twinning the human’s state and behavior to also twinning its looks, thus improving the user’s identification with their HDT, for example, through body ownership \cite{genay2021, waltemate2018}. Several authors suggest using human modeling tools such as Siemens Jack \cite{SiemensJack}, SantosHuman \cite{SantosHuman}, and ObEN \cite{Oben}, which allow modeling the human entity and graphical simulations \cite{kerckhove2021, demirel2021}. In addition, the adoption of tools from other domains, such as gaming engines, could allow for avatars with even advanced detail and customizability \cite{birk2018}. This way, user avatars could improve the HDT’s perception by the user as a visual twin of themself, allowing for improved understandability, and even for advanced cognitive mechanisms such as immersion and self-reflection \cite{hooi2012}.\\
Many applications, however, might not provide graphical or gamified user interfaces. Instead, HDTs could be embodied in other aspects of the system, such as social robots or conversational agents \cite{kerckhove2021, maeyer2020}. This way, users would not feel body ownership but could potentially see the HDT as a virtual friend or companion \cite{maeyer2020}. \cite{abeydeera2019}, for example, proposes supporting the user’s well-being and everyday life by integrating their HDT into a smart-mirror system that visualizes suggestions directly on the user’s mirror image. Therefore, the embodiment of HDTs is not limited to human-like avatars, but poses a multifaceted and yet-to-be-explored concept whose realization should be tailored to the specific use-case.\\
Fourth, the HDT’s user interface must provide \textbf{simplicity and intuitiveness}. Depending on the use-case, HDT applications typically entail haptic (e.g., through exoskeletons \cite{pizzolato2019}), acoustic (e.g., through verbal interaction \cite{kerckhove2021}), or visual feedback. Particularly the latter poses the question of how to best communicate relevant information to the user. As already discussed, the embodiment of HDTs could be a promising approach to increase the HDTs graspability. Combining avatars with advanced visualization techniques such as animations could be an even more powerful way to visualize human attributes intuitively \cite{reipschläger2022}. Additionally, the choice of the medium on which to implement the user interface can significantly affect the intuitiveness, as well as the user’s perception of the HDT. While web-based interfaces can provide broad accessibility through compatibility with common web browsers \cite{han2021}, AR/VR technology could allow for intuitive data visualization and immersion, as well as for new possibilities of interacting with the own HDT, with virtual environments, or even collaboratively with HDTs of other users \cite{bravo2020, waltemate2018, schäfer2022}. Therefore, the choice of user interface as well as the concrete feedback design significantly affects the user’s perception of the HDT depending on its simplicity and intuitiveness or even triggers deeper cognitive mechanisms, such as immersion.\\
Together, providing identifiability and traceability, credibility and explainability, and simplicity and intuitiveness can help to increase the user’s trust in the system through enhanced transparency and thus improve the system’s effectiveness.\\\\

\subsubsection{Ethical Considerations}
\label{Section5.3.4}
Three key design considerations emerge regarding the ethics of HDTs.\\
First, HDTs must ensure \textbf{data security and privacy}. Table \ref{Table2} highlights significant concerns for data privacy and security amongst users, particularly given the high sensitivity of personal data. For this reason, various regulations (a comprehensive overview about typical regulations can be found in \cite{aljeraisy2021}) exist to ensure the privacy and security of the user's data, that must be taken into account in the development of HDTs. Apart from the security measures described in Section \ref{Section2.3}, data anonymization could—to some degree—improve data security and privacy and thus help to fulfil the requirements posed by these regulations. \cite{Angulo2020}, for example, suggests using General Adversarial Networks to generate “fake” data similar to the original user data. In addition to contributing to data security by depersonalizing the data, the generated data could further be used to train prediction models. These would therefore be independent of actual user data, allowing the complete withdrawal of personal data as required by regulation without affecting the HDTs overall performance \cite{Angulo2020}. However, this approach could also affect the HDT’s accuracy and reliability, depending on the quality of the generated data. In addition, \cite{petrova2020} suggests data separation methods to improve data security. Therefore, the user profile can be stored separately from the corresponding user data, for instance, using blockchain technology, resulting in a distribution of data, thus reducing the risk of data theft \cite{petrova2020, suhail2022}. However, this could lead to complex and resource-expensive systems, thus decreasing the HDTs transparency. Further guidance on security measures can be found, for instance, in \cite{kayan2022} and \cite{waheed2020}. Therefore, data anonymization and separation can help increase data security and privacy and promote regulation conformity but could also affect other aspects of the HDT.\\
Second, in addition to protecting user-related data, HDTs must also ensure \textbf{safe interaction} with the users themselves, both regarding direct interaction through physical feedback and indirect interaction through decision support. One way to achieve this is the already mentioned use of high-quality and consistent data and the validation of the reliability of prediction modules. In addition, adaptability of HDTs again plays a major role in this context, as it ensures a continuous adaptation of the prediction modules to dynamically changing contexts. However, such (autonomous) evolution of the HDT may also lead to changes in reliability, which would theoretically require the system to be revalidated after each change. Therefore, \cite{lutze2019} suggests the use of self-validation techniques that ensure the reliability of predictions and decision support after each change. Regarding the safety of direct physical feedback, \cite{pizzolato2021} also suggests using tailored physiological models to set appropriate boundary conditions for the feedback. Thus, frequent (self-)validations and careful determination of boundary conditions can increase the safety of both direct and indirect feedback.\\
Third, to maximize its benefit and prevent social bias and injustice, HDT technology—in the long term—should be \textbf{equally distributed}. While this ethical challenge and its concrete extent is highly complex and not yet answered, it correlates with several other previously discussed considerations. A simple and intuitive user interface, for example, would significantly reduce the expert knowledge required to handle the HDT. Further, standardization and modularity would reduce the HDT’s complexity and costs, thus increasing its affordability. In addition, \cite{martinez2019} and \cite{diaz2021} propose first approaches for HDTs that primarily operate on edge devices, reducing their dependency on complex IT infrastructure. Thus, although some of the discussed considerations could represent initial steps towards equally distributed HDT technology, this challenge—due to its complexity—is still far from being answered sufficiently and thus needs further elaboration.\\
Together, ensuring data security and privacy, user safety, as well as promoting equal distribution of the HDT system  provide key steps to addressing ethical challenges in HDTs, yet also highlight the need for significant further work in this area. Table \ref{Table3} summarizes the design considerations discussed in this section.\\\\

{\centering
\begin{longtable}{p{0.3\textwidth} p{0.6\textwidth}}
 \caption{11 key considerations in Human Digital Twin (HDT) design.}
 \label{Table3}\\
    \toprule
    Consideration     & Description \\
    \midrule
    \multicolumn{2}{l}{\textit{Data Collection and Processing}} \\
    \midrule
    Fidelity and Accuracy & HDTs must represent the human entity with sufficient accuracy for the particular use case. Combining modeling and sensor data fusion can help to address complexity-related challenges and provide sufficient fidelity and accuracy within the HDT.\\
    Sensitivity and Robustness  & HDTs must be robust and able to handle inconsistent data to provide reliable feedback. Including data processing capabilities can compensate for inconsistent and low-quality data, improving the overall reliability, and thus the trustworthiness of HDTs. \\
    \midrule
    \multicolumn{2}{l}{\textit{Flexibility and Interconnectivity}} \\
    \midrule
    Interconnectivity & HDTs must be able to communicate with systems and DTs in their environment. Implementing standardized communication protocols can help to integrate HDTs seamlessly into DT networks.\\
    Adaptability  & HDTs must be able to adjust to changes within the human entity, its environment, and the application context. Using modular design helps to adapt parts of the HDT architecture without affecting other aspects, whereas including continuous learning techniques can help to dynamically adjust the HDT’s decision-making to new data. \\
    \midrule
    \multicolumn{2}{l}{\textit{Transparency and Trust}} \\
    \midrule
    Identifiability and Traceability & HDTs must be clearly assignable to their human entity and provide information about the history of significant changes. Assigning unique identification numbers can help to prevent confusing HDTs of different entities and including versioning allows for checking the current update status.\\
    Credibility and Explainability  & HDTs must provide understandable and reliable feedback to promote the user’s trust in the system. Integrating professionals and domain knowledge can help to address complexity challenges in the development process, whereas testing the resulting system using profound validation metrics can help prove the HDTs reliability.  \\
    Embodiment  & HDTs can be embodied in visualizations or other system components to improve understandability, intuition, and trust. Avatars, social robots, or virtual assistants are promising examples of system components that can be connected to the HDT to interact with the human in a trustworthy and intuitive way.\\
    Simplicity and Intuitiveness & HDTs must provide simple and intuitive user interfaces when directly interacting with users to prevent confusion and frustration. Feedback can be haptic, acoustic, and/or visual and can be combined with web-based interfaces for easy accessibility or virtual/augmented reality interfaces for advanced immersion. \\
    \midrule
    \multicolumn{2}{l}{\textit{Ethical considerations}} \\
    \midrule
    Data security and privacy & HDTs must ensure data security and privacy due to the high sensitivity of human-specific data. Including profound high-end security measures, ensuring regulation conformity, and following a user-oriented development process can help to build trust and facilitate handling the challenges arising from data-ownership questions. \\
    Safe Interaction  & HDTs must ensure reliability and safety regarding their direct and indirect feedback, as incorrect feedback could pose serious risks to the user's well-being. Continuous (self-) validation and appropriate boundary conditions can help increase the generated feedback's reliability and safety.  \\
    Equally distributed  & HDTs must be accessible to all countries and social groups. Considering how specific social groups interact with the HDTs and ensuring equal representation within the underlying data could help to prevent the promotion of social inequalities and bias.\\
    \bottomrule
\end{longtable}}

The considerations in Table \ref{Table3} provide guidance on how to design HDTs, responding to the challenges identified in Table \ref{Table2}. Within the reviewed HDT literature, only a select few papers provide similar considerations. However, they either refer to DTs in general (e.g., \cite{barricelli2019}) and thus do not fully address the human-specific factors—especially perception, trust, and ethics issues—or are limited to specific applications (e.g., \cite{rivera2019}). Within the broader literature on human-centric system design, several important general design considerations are proposed, including user support through personalization and guidance \cite{oinas-kukkonen2009}, support of social interactivity \cite{oinas-kukkonen2009}, and safe and intelligent user-system interaction \cite{zarte2020}. However, these considerations do not account for the specific characteristics of HDTs that substantially impact how such considerations might be applied, as outlined in Section \ref{Section4}. Therefore, our proposed considerations address a critical need in the literature by taking both human factors and DT-specific context into account.

\subsection{Limitations}
\label{Section5.4}
Despite their informative value and grounding in cross-domain HDT literature, the results discussed in this paper should be considered in light of two main limitations.\\
First, the proposed definition (Figure \ref{Fig.7}), challenges (Table \ref{Table2}), and design considerations (Table \ref{Table3}), are necessarily generalized to capture a cross-domain perspective, while also remaining valid for all currently existing HDT applications. This necessitates a degree of abstraction from specific applications, as well as the acknowledgment that this is an emerging field, and hence demands careful consideration when implementing HDTs in practice, especially in new domains. For example, although our definition highlights the importance of ongoing synchronization with the physical entity and the provision of direct or indirect feedback, it does not define the (minimal) extent to which these attributes have to be manifested in the HDT. Thus, while we provide an important foundational step towards formalizing the concept of HDTs and distinguishing it from related technologies, we also highlight the need for further work in examining how these general characteristics should be adapted to specific HDT implementations.\\
Second, the results must also be considered formative due to the preliminary and emerging nature of the underlying literature. As indicated in Figure \ref{Fig.4}, the research field of HDTs is highly recent and to-date encompasses only a few papers that propose concrete implementations of HDTs. Accordingly, the insights elaborated in this paper are to a large extent based on general discussions, frameworks, and knowledge transfer from related fields such as traditional DTs. Further, due to the early stage of the research field, some important considerations are still unexplored. For instance, the integration of the user's behavioral aspects such as intention, social behavior, routines, and habits—which could complement the HDT with important contextual information—are barely addressed in the current literature. Thus, while this paper distils central aspects in the design of HDTs—which are fundamental and will likely remain crucial also in the future—their concrete manifestation might change and additional considerations might emerge as the field evolves, particularly in terms of implementations of HDTs.\\
Despite these limitations, the results developed in this paper play an important role in concretizing the emerging field of HDTs by distilling general considerations concernnig the design of HDTs. As such, this paper should be seen as part of the evolution process, providing a guideline that encourages and facilitates the development of future HDT applications.

\subsection{Implications and Future Work}
\label{Section5.5}
\begin{figure} [H]
  \centering
  \includegraphics[scale=0.57]{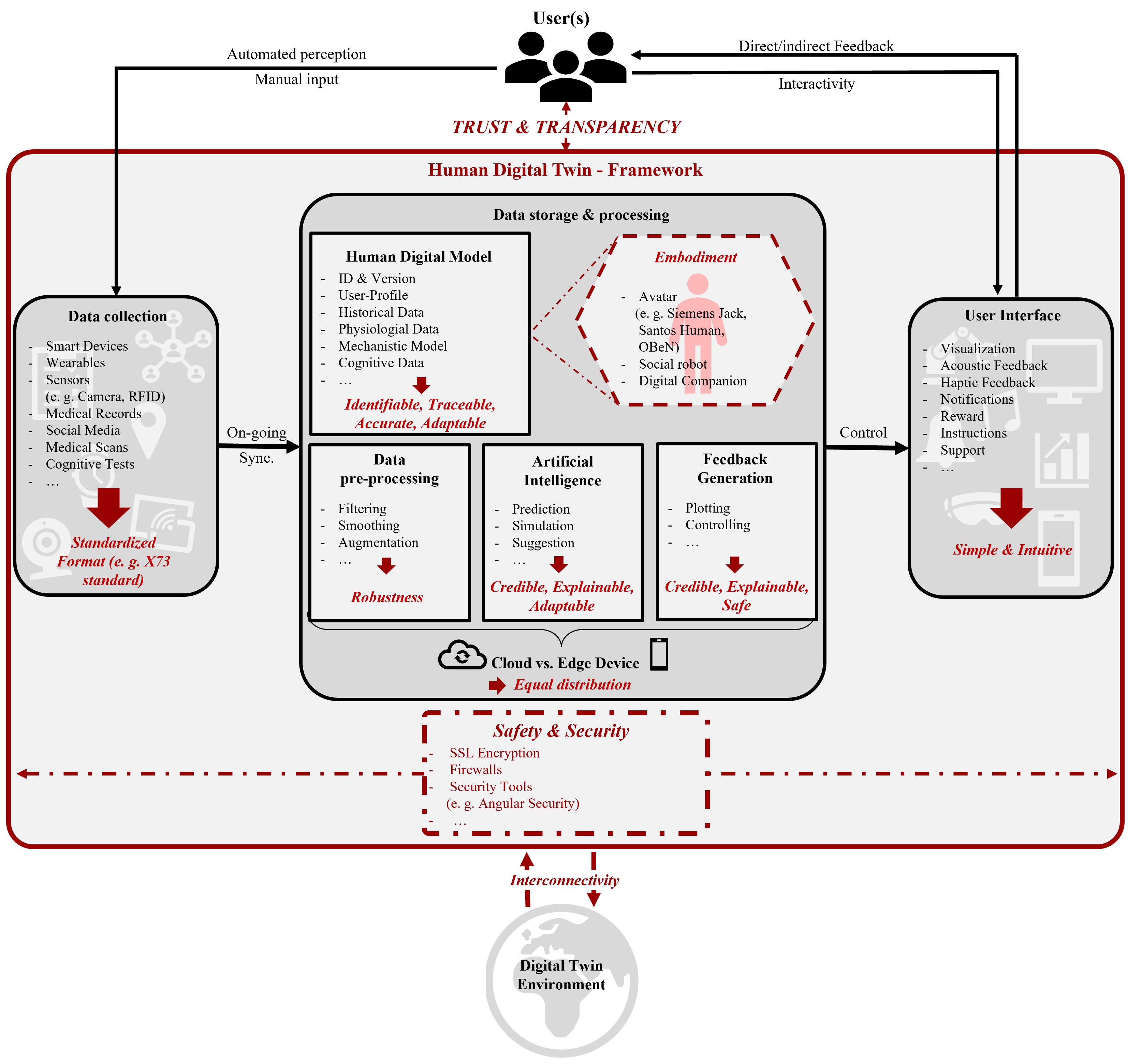}
  \caption{General Human Digital Twin-Framework that extends the conventional Digital Twin framework with the elaborated human factors and considerations – denoted in cursive letters.}
  \label{Fig.8}
\end{figure}

Bringing our findings together (definition and Figure \ref{Fig.7}, challenges in Table \ref{Table2}, and design considerations in Table \ref{Table3}), we are able to offer a conceptual framework for HDTs, shown in Figure 8.\\
Comparing Figure \ref{Fig.8} with the traditional DT framework in Figure \ref{Fig.2} reveals key similarities and differences between HDTs and DTs. Here, common elements include the three fundamental layers, perception of the physical world with sensor data, and translation of aggregated data into corresponding feedback. However, several elements in the HDT framework take on an expanded or significantly modified role. For example, while traditional DTs typically use integrated sensors, HDTs mostly rely on external, intrusive sensors or manual input to perceive the human’s state. Further, HDTs introduce several new elements. For example, human factors, such as trust, are not found in machines or processes but are crucial for the user’s perception and acceptance of HDTs. Thus, the framework extends the already well-established concept of the traditional DT by adapting key features to the human-oriented context and adding new considerations emerging from human factors. This conceptualization and the underlying findings provide three main implications.\\
First, we analyzed the current HDT literature for fundamental characteristics and proposed an initial general, cross-domain definition of HDTs. To the authors’ knowledge, this definition is the first explicit approach to defining HDTs as a stand-alone cross-domain concept and clearly differentiating it from other related concepts, such as digital models or shadows. The definition reveals the large potential variety in possible HDTs due to the various aspects and abstraction levels of the human entity, but at the same time exposes the current infeasibility of creating an all-encompassing, complete HDT given this variety and complexity. Further, the definition is particularly sensitive regarding the on-going synchronization between HDT and the human entity and the provision of feedback. However, the limited guidance on their concrete realization challenges designers and developers to identify the optimal regularity and frequency of synchronization, as well as the ideal kind of feedback considering the underlying use-case.\\
Second, we distilled major challenges that arise from the HDT concept. These challenges extend those identified in prior works in a number of important areas, including perception, trust, and ethics. Further, the identified challenges show that the technical development of HDTs and the user-specific factors cannot be dissociated. In fact, shattered trust of the user in the HDT can result in system failure, regardless of its technological viability. Thus, considering user-specific requirements must play a decisive role in the development of HDTs from the very beginning.\\
Third, we derived a number of key design considerations. To the author’s knowledge, these considerations are the first of their kind and provide future research with guidance on how to face the HDT-related challenges. However, as discussed in the previous section, although Table \ref{Table3} highlights essential design considerations, such as the prioritization of data security and privacy for user protection, these do not sufficiently address the high complexity and multifaceted nature of HDTs, particularly regarding emerging ethical challenges and behavioral aspects. Thus, further work needs to be carried out to identify further ethical challenges as well as concrete design considerations addressing them. Therefore, these implications lead to a number of important directions for future work. \\
First, in the near future, more HDT applications need to be developed within the described domains by translating the proposed HDT framework into concrete implementations. This would form an essential basis for tackling domain-related questions surrounding, for example, how HDTs could be embodied or represented to increase the effectiveness of the HDT application and the trust in it. In addition, opportunities for the adaptation of the HDT concept in new domains should be explored in order to better understand and critically develop the generalizability of HDT characteristics and design guidelines.\\
Second, once a larger pool of HDT implementations is available, domain-specific reviews need to be conducted to identify the associated additional requirements and challenges to complement the definition and design considerations proposed in this paper. Particularly, considerations that address ethical challenges and the integration of behavioral aspects need to be further explored. These are important areas of development addressed in only few HDT applications to date, thus raising the need to bring together insights as they emerge from new research.\\
Third, in the long term, standards need to be established particularly concerning data communication and security. In addition, the ethical challenges and societal impacts that arise from HDT applications need to be explored further to develop appropriate detailed guidelines. This is critical to ensuring the continued ethical and social acceptability of HDTs and bringing these challenging discussions to the forefront of an emerging field, which has to date generally neglected such issues.\\

\section{Conclusion}
\label{Section6}
Human Digital Twins (HDTs) are a fast-emerging technology with significant potential in fields ranging from healthcare to sports. However, there is a critical need to clarify what HDTs are, how they have evolved and been used, and how they can be designed. Thus, in this paper, we set out to explore the HDT concept, building on a first systematic review of 2049 papers and a focused review of 70 papers related to human digital twins (HDT) across various domains. Based on this, we provide a first cross-domain definition of HDTs (Section \ref{Section5.1}) and derive 14 major challenges (Table \ref{Table2}) and eleven key considerations (Table \ref{Table3}) impacting their design and development. This understanding of HDTs and how they differ from traditional Digital Twins is encapsulated in Figure \ref{Fig.8}.\\
Together these results provide the basis for a number of theoretical and practical implications surrounding the research and implementation of HDTs across different domains. Not least of these was that, due to the emerging nature of the discussion surrounding HDTs, there is a particular need for further study of their implementation and design, as well as on additional ethical challenges and corresponding design considerations. As such, in addition to generating a common understanding based on current research, this paper also serves as a call to action for further research in this area.

\section*{Acknowledgments}
This work has received funding from the Horizon 2020 research and innovation program of the European Union under grant agreement no. 871767 of the project ReHyb: Rehabilitation based on hybrid neuroprosthesis.

\bibliographystyle{unsrt}  
\bibliography{references}  

\section*{Appendix}
{\centering
\begin{longtable}{p{0.05\textwidth} p{0.05\textwidth} p{0.05\textwidth} p{0.55\textwidth} p{0.05\textwidth} p{0.05\textwidth} p{0.05\textwidth} p{0.05\textwidth}}
 \caption{Overview of the reviewed papers and their contribution to the analytical results (Section \ref{Section4}; Def. = “Definition”, Techn. = “Technical Conceptualization”, P\&A = “Perception and Acceptance”).  “\checkmark” indicates strong focus on the corresponding theme, “(\checkmark)” indicates mentioning an indirect derivability, and “–“ indicates no addressing of the corresponding theme.}
 \label{Table4}\\
    \toprule
    Ref.     & Year     & Type     & Description     & Def.     & Techn.     & P\&A     & Ethics \\
    \midrule
    \multicolumn{8}{l}{Healthcare} \\
    \midrule
    \cite{rivera2019} & 2019 & C & Framework for Human Digital Twins (HDTs) in precision medicine. & \checkmark & \checkmark & \checkmark & (\checkmark)\\
    
    \cite{Schade2019} & 2019 & J & HDT of cancer patients for individualized drug dosage plan in cancer treatment. & – & \checkmark & – & –\\

    \cite{Goodwin2020} & 2019 & J & Development of a metabolic Digital Twin (DT) to predict the blood glucose level responses of humans for determining an insulin injection policy. & (\checkmark) & \checkmark & \checkmark & –\\

    \cite{pizzolato2019} & 2019 & J & Development of a neuromusculoskeletal model of patients for improved therapy after spinal court injuries. & (\checkmark) & \checkmark & \checkmark & \checkmark\\

    \cite{nonnemann2019} & 2019 & C & Introduction of an IoT system for monitoring vital and administrative data in the digital twin of an intensive care unit. & - & (\checkmark) & (\checkmark) & -\\

    \cite{Lauzeral2019} & 2019 & J & HDT in form of a mechanical model of the human liver with patient-specific geometry and real-time capabilities for predicting the liver's deformation during breathing. & (\checkmark) & \checkmark & - & -\\

    \cite{martinez2019} & 2019 & C & HDT of human heart based on sensor data from ECG, medical records, and others for detecting heart disease and stroke. & (\checkmark) & \checkmark & (\checkmark) & -\\

    \cite{Comito2019} & 2019 & C & Framework for clinical decision support using basic health records and social media data. & \checkmark & \checkmark & - & -\\

    \cite{alcaraz2019} & 2019 & C & Introduction of a Memory Polynomial Model for reducing the amount of IMUs required to build an HDT for gait assessment. & - & \checkmark & (\checkmark) & -\\

    \cite{lutze2019} & 2019 & C & Introduction of a new approach for unlearning patient-specific data in HDTs in compliance with the new EU-GDPR regulations. & (\checkmark) & \checkmark & \checkmark & (\checkmark)\\

    \cite{kenzierskyj2019} & 2019 & B & Discussion of issues, concerns, and ethical implications of Big Data, Data Harvesting, and DTs in healthcare. & (\checkmark) & (\checkmark) & \checkmark & \checkmark\\

    \cite{Liu2019} & 2019 & J & Cloud-Based HDT Framework that uses data from medical records, wearable devices etc. for the Elderly Healthcare. & (\checkmark) & \checkmark & (\checkmark) & (\checkmark)\\

    \cite{Angulo2020} & 2020 & C & General decision-support HDT framework in healthcare that uses a data-anonymization method based on GANs. & (\checkmark) & \checkmark & (\checkmark) & \checkmark\\

    \cite{Mohapatra2020} & 2020 & J & Review about the design of (H)DT frameworks in healthcare. & (\checkmark) & \checkmark & (\checkmark) & (\checkmark)\\

    \cite{shamanna2020_1} & 2020 & J & Assessment of changes in ambulatory glucose profile metrics of patients in the course of a digital twin-driven precise nutrition program. & (\checkmark) & (\checkmark) & - & -\\

    \cite{shamanna2020_1} & 2020 & J & Assessment of changes in ambulatory glucose profile metrics of patients in the course of a digital twin-driven precise nutrition program. & (\checkmark) & (\checkmark) & - & -\\
    
    \cite{Erol2020_2} & 2020 & C & Review about DTs in the healthcare domain and their development. & (\checkmark) & (\checkmark) & (\checkmark) & -\\
    
    \cite{petrova2020} & 2020 & C & HDT that evaluates the cognitive state of the patients using cognitive test results in order to determine behavior change in patients with Multiple Sclerosis. & (\checkmark) & \checkmark & (\checkmark) & (\checkmark)\\
        
    \cite{rodriguez-aguilar2020} & 2020 & J & Development of a framework for a digital public health emergency system for managing human, financial and material resources as well as applying multi-paradigm simulation. & (\checkmark) & (\checkmark) & - & -\\
           
    \cite{corral2020} & 2020 & J & Review about HDTs in human cardiovascular system modeling. & (\checkmark) & \checkmark & \checkmark & \checkmark\\
               
    \cite{Bagaria2020} & 2020 & B & Review about current literature on DTs for health and wellbeing. & (\checkmark) & (\checkmark) & (\checkmark) & (\checkmark)\\
               
    \cite{derungs2020} & 2020 & J & Introduction of three new digital biomarkers for gait assessment of stroke patients. & (\checkmark) & (\checkmark) & - & -\\
           
    \cite{Calderita2020} & 2020 & J & CORTEX-Digital Twin-based cyber-physical system environment for ambient assistant living. & - & (\checkmark) & - & (\checkmark)\\
              
    \cite{maeyer2020} & 2020 & C & Survey about how HDT should be designed for wellbeing and elderly care from the patient and professional perspective. & (\checkmark) & (\checkmark) & \checkmark & \checkmark\\
                 
    \cite{Laamarti2020} & 2020 & J & ISO/IEEE 11073 standardized (H)DT framework for personal well-being and smart cities that uses data from e. g. personal health devices while including ISO/IEEE 11073 support. & (\checkmark) & \checkmark & \checkmark & (\checkmark)\\
               
    \cite{Lutze2020} & 2020 & C & Health domain-specific DT framework and its requirements with special focus on EU and FDA regulations. & (\checkmark) & \checkmark & \checkmark & (\checkmark)\\
                  
    \cite{boer2020} & 2020 & J & Discussion on how HDTs affect the human’s perception of their body. & \checkmark & (\checkmark) & (\checkmark) & \checkmark\\
                  
    \cite{gkouskou2020_1} & 2020 & J & Discussion on how to create a digital virtual patient twin incorporating various human factors for precise nutrition. & (\checkmark) & \checkmark & - & (\checkmark)\\
                      
    \cite{croatti2020} & 2020 & J & Review about agent-based DTs in healthcare on the use-case of trauma management. & (\checkmark) & (\checkmark) & - & -\\
                  
    \cite{marques2020} & 2020 & J & Evaluation of the effect of the sensor position and the sitting position of the patient on the accuracy of the knee range of movement (ROM) data displayed by an e-rehabilitation system. & - & - & (\checkmark) & -\\
                 
    \cite{friebe2020} & 2020 & C & Review about innovation needs in the biomedical engineering curriculum. & - & - & (\checkmark) & (\checkmark)\\
                 
    \cite{boulos2021} & 2021 & J & Review about HDTs in health and medicine. & (\checkmark) & \checkmark & \checkmark & \checkmark\\
                     
    \cite{wickramasinghe2021} & 2021 & J & Review about HDT in healthcare and their application in cancer treatment. & (\checkmark) & \checkmark & \checkmark & (\checkmark)\\
                     
    \cite{braun2021} & 2021 & J & Review about HDTs and their scope with respect to ethical aspects. & (\checkmark) & \checkmark & \checkmark & \checkmark\\
                     
    \cite{kerckhove2021} & 2021 & J & Discussion on cognitive functionalities and ethical considerations in personal HDTs. & \checkmark & (\checkmark) & (\checkmark) & \checkmark\\
                   
    \cite{popa2021} & 2021 & J & Discussion/Survey about socio-ethical benefits and risks of DTs in healthcare & (\checkmark) & (\checkmark) & (\checkmark) & \checkmark\\
                     
    \cite{mourtzis2021} & 2021 & C & HDT Framework for visualizing data from magnetic resonance imaging and processing and analyzing patient data using AI. & - & (\checkmark) & (\checkmark) & -\\
                       
    \cite{shamanna2021} & 2021 & J & Cohort study about how HDT-based treatment on type-2-diabetes on body mass index, blood pressure, and the use of antihypertensive medications. & (\checkmark) & (\checkmark) & - & -\\
                         
    \cite{dillenseger2021} & 2021 & J & Review about enabling technologies in the field of digital biomarkers for creating an HDT for multiple sclerosis patients. & \checkmark & (\checkmark) & (\checkmark) & (\checkmark)\\
                           
    \cite{voigt2021} & 2021 & J & Review about DTs in multiple sclerosis treatment. & \checkmark & \checkmark & \checkmark & \checkmark\\
                          
    \cite{pizzolato2021} & 2021 & J & Review about efficacious non-invasive therapies that are increasingly used to restore function in people with chronic spinal cord injuries. & (\checkmark) & (\checkmark) & (\checkmark) & -\\
                           
    \cite{firouzi2021} & 2021 & J & Review and discussion about how current technologies could contribute to fighting the COVID-19 pandemics. & - & (\checkmark) & \checkmark & \checkmark\\
                           
    \cite{shaked2021} & 2021 & J & Development of a data-driven tool for improved hospital organization and monitoring. & (\checkmark) & (\checkmark) & (\checkmark) & -\\
                          
    \cite{elayan2021} & 2021 & J & Proposal of an HDT framework for healthcare application on the use-case of anomaly detection of the human heart. & (\checkmark) & \checkmark & \checkmark & \checkmark\\
                           
    \cite{chase2021} & 2021 & C & Review about DTs in medical critical care units. & (\checkmark) & \checkmark & (\checkmark) & (\checkmark)\\
                           
    \cite{dang2021} & 2021 & J & Review about challenges and opportunities in building an actionable AI model for education and decision support in neurocritical care. & (\checkmark) & - & \checkmark & -\\
                          
    \cite{shamanna2021_2} & 2021 & J & Type 2 diabetes reversal with digital twin technology-enabled precision nutrition and staging of reversal. & (\checkmark) & - & - & -\\
                         
    \cite{chakshu2021} & 2021 & J & Creating an HDT of the human vascular system using inverse analysis based on recurrent neural networks. & (\checkmark) & \checkmark & (\checkmark) & -\\
                         
    \cite{heininger2021} & 2021 & C & Design of autonomous DT-based cyber-physical systems for home healthcare. & (\checkmark) & (\checkmark) & - & (\checkmark)\\
                         
    \cite{ricci2021} & 2021 & J & Review about DTs in health organization. & - & (\checkmark) & - & -\\
                         
    \cite{zheng2021} & 2021 & C & Proposal of a privacy-preserving cloud-based DT framework in healthcare. & \checkmark & (\checkmark) & - & \checkmark\\
    
    \midrule
    \multicolumn{8}{l}{Industry (incl. oil/gas industry, construction and pharma industry)} \\
    \midrule
                         
    \cite{Savur2019} & 2019 & C & Introduction of an experimentation platform for human-robot collaboration. & (\checkmark) & (\checkmark) & (\checkmark) & (\checkmark)\\
                        
    \cite{lv2021} & 2021 & J & Proposal of a framework for human-robot-collaborative assembly based on DT. & (\checkmark) & \checkmark & - & (\checkmark)\\
                        
    \cite{wangT2021} & 2021 & J & Proposal of a DT framework for human-machine collaboration based on visual question answering. & - & (\checkmark) & (\checkmark) & -\\
                        
    \cite{wangZ2021} & 2021 & J & DT of an Injection Molding Smart Factory. & - & (\checkmark) & - & -\\
                       
    \cite{wanasinghe2021} & 2021 & J & Human-centric digital transformational framework for the oil and gas industry to deploy existing digital
technologies to enhance their workers’ health, safety, and working conditions. & - & (\checkmark) & \checkmark & \checkmark\\
                  
    \cite{montini2021} & 2021 & C & HDT that monitors human workers for improved well-being and process performance in the industrial domain. & (\checkmark) & \checkmark & - & -\\
                  
    \cite{tzavara2021} & 2021 & C & Review about human detection, human task monitoring, and digital twin integration in human-robot-collaborative industry. & (\checkmark) & (\checkmark) & - & \checkmark\\
                  
    \cite{canzoneri2021} & 2021 & B & Review about DTs in the bio-pharma industry. & (\checkmark) & \checkmark & \checkmark & \checkmark\\
                 
    \cite{demirel2021} & 2021 & J & Proposal of a human factors engineering approach for designing products while considering human factors and models. & - & (\checkmark) & - & (\checkmark)\\
                 
    \cite{hou2021} & 2021 & J & Review of DTs in construction work safety. & - & (\checkmark) & - & \checkmark\\
                 
    \cite{akanmu2021} & 2021 & J & Review about opportunities of cyber-physical systems in the construction industry. & - & (\checkmark) & (\checkmark) & (\checkmark)\\

    \midrule
    \multicolumn{8}{l}{Sports)} \\
    \midrule
                 
    \cite{abeydeera2019} & 2019 & C & Smart mirror system that supports the user in their everyday life well-being by capturing their optical appearance, detecting diseases, and suggesting improvements. & - & \checkmark & \checkmark & -\\
                
    \cite{Hafez2019} & 2019 & C & HDT to align human objectives with surrounding smart devices. & (\checkmark) & \checkmark & - & (\checkmark)\\
    
    \cite{diaz2020} & 2020 & J & Review about HDT in coaching and wellbeing, the requirements and challenges, and introduction of a general HDT coaching concept. & (\checkmark) & \checkmark & \checkmark & \checkmark\\
                
    \cite{barricelli2020} & 2020 & J & Athlete HDT-based application to predict their physical training performance during the day based on their behavior scheme to provide decision support for trainers and coaches. & (\checkmark) & (\checkmark) & \checkmark & -\\
                
    \cite{diaz2021} & 2021 & M & Edge-running DT-based coaching application for physical activities during COVID-19. & (\checkmark) & (\checkmark) & (\checkmark) & (\checkmark)\\

    \midrule
    \multicolumn{8}{l}{Others)} \\
    \midrule

    \cite{barricelli2019} & 2019 & J & Review about DTs and their design in several domains. & (\checkmark) & \checkmark & \checkmark & \checkmark\\
    
    \cite{Fuller2020} & 2020 & J & Review about DTs in different domains. & (\checkmark) & \checkmark & \checkmark & \checkmark\\
    
    \cite{Erol2020_1} & 2020 & C & Review about the current state of DTs in different domains. & (\checkmark) & - & - & -\\
    
    \cite{wu2021} & 2021 & J & Review about DT networks and their key technologies, applications, challenges, and trends. & (\checkmark) & \checkmark & \checkmark & (\checkmark)\\

    \bottomrule
\end{longtable}}

\end{document}